\documentclass[aps,pra,twocolumn,showpacs,amsmath,amssymb,superscriptaddress]{revtex4-1}
\usepackage[english]{babel}
\usepackage{amsmath,amsfonts,bm,graphicx,verbatim,mathrsfs,color}
\usepackage[colorlinks=true,citecolor=red]{hyperref} 

\begin{document}

\title{Extreme reductions of entropy in an electronic double dot}

\author{Shilpi Singh}
\affiliation{Low Temperature Laboratory, Department of Applied Physics, Aalto University, 00076 Aalto, Finland}

\author{ \'Edgar Rold\'an}
\affiliation{ICTP - The Abdus Salam International Centre for Theoretical Physics, Strada Costiera 11, 34151, Trieste, Italy}
\affiliation{Max Planck Institute for the Physics of Complex Systems, N\"othnitzer Stra{\ss}e 38, 01187 Dresden, Germany}
\affiliation{Center for Advancing Electronics Dresden cfAED, 01062 Dresden, Germany}

\author{Izaak Neri}
\affiliation{Max Planck Institute for the Physics of Complex Systems, N\"othnitzer Stra{\ss}e 38, 01187 Dresden, Germany}
\affiliation{Mathematics Department, King's College London, Strand, London WC2R 2LS, UK}

\author{Ivan M. Khaymovich}
\affiliation{Max Planck Institute for the Physics of Complex Systems, N\"othnitzer Stra{\ss}e 38, 01187 Dresden, Germany}
\affiliation{Institute for Physics of Microstructures, Russian Academy of Sciences, 603950 Nizhny Novgorod, GSP-105, Russia}

\author{Dmitry~S.~Golubev}
\affiliation{Low Temperature Laboratory, Department of Applied Physics, Aalto University, 00076 Aalto, Finland}

\author{Ville F. Maisi}
\affiliation{NanoLund and the Department of Physics, Lund University, Box 118, S-22100 Lund, Sweden}

\author{Joonas T. Peltonen}
\affiliation{Low Temperature Laboratory, Department of Applied Physics, Aalto University, 00076 Aalto, Finland}

\author{Frank J\"ulicher}
\affiliation{Max Planck Institute for the Physics of Complex Systems, N\"othnitzer Stra{\ss}e 38, 01187 Dresden, Germany}

\author{Jukka P. Pekola}
\affiliation{Low Temperature Laboratory, Department of Applied Physics, Aalto University, 00076 Aalto, Finland}



\begin{abstract}
We experimentally study negative fluctuations of stochastic entropy production in an electronic double dot operating in nonequilibrium steady-state conditions.
We record millions of random electron tunneling events at different bias points, thus collecting extensive statistics.
We show that for all bias voltages the experimental average values of the minima of stochastic entropy production lie above  $-k_{\rm B}$,
where $k_{\rm B}$ is the Boltzmann constant, in agreement with recent theoretical predictions for nonequilibrium steady states.
Furthermore, we also demonstrate that the experimental cumulative distribution of the entropy production minima is bounded, at all times and for all bias voltages, by a universal
expression predicted by the theory.
We also extend our theory by deriving a general bound for the average value of the maximum heat
absorbed by a mesoscopic system from the environment and compare this result with experimental data.
Finally, we show by numerical simulations that these results are not necessarily valid under non-stationary conditions. 
\end{abstract}
\maketitle

%

\section{Introduction}
\label{s1}

According to the second law of thermodynamics, the entropy production, given by the entropy change of a macroscopic system plus the entropy change of its environment, can only grow in time.
However, mesoscopic systems can sometimes move against the
tide due to fluctuations.
As a result, the entropy production in small systems, such as a single-electron box, fluctuates and, while growing on average, can decrease during short time intervals when an
electron tunnels in the direction opposite to the electric force~\cite{RevModPhys.81.1665,sekimoto2010stochastic,jarzynski2011equalities,seifert2012stochastic,esposito2015ensemble,Campisi2015NJP}.
The ratio of the probabilities for the entropy production to take positive or negative values is determined by fluctuation relations derived in theory and successfully demonstrated
in a plethora of experimental setups of different nature (e.g. DNA molecules, colloidal systems, RC circuits and single-electron boxes~\cite{bustamante2005nonequilibrium,OPS_PRL.109.180601,koski644,pekola2015towards,khaymovich2015multifractality,ciliberto2017experiments,gavrilov2017direct,martinez2017colloidal}).
Recently, it was shown that applying an appropriate periodic drive to a mesoscopic system in combination
with a feedback-control, such as in mesoscopic Maxwell-demon experiments~\cite{toyabe2010experimental,berut2012experimental,jun2014high,koski2014experimental,roldan2014universal,pekola2018thermodynamics},
one can even achieve the reduction in the average value of its entropy of the order of $k_{\rm B}$ per one cycle,
where $k_{\rm B}$ is Boltzmann's constant.

\begin{figure}[ht!]
 \includegraphics[width= 0.8 \linewidth]{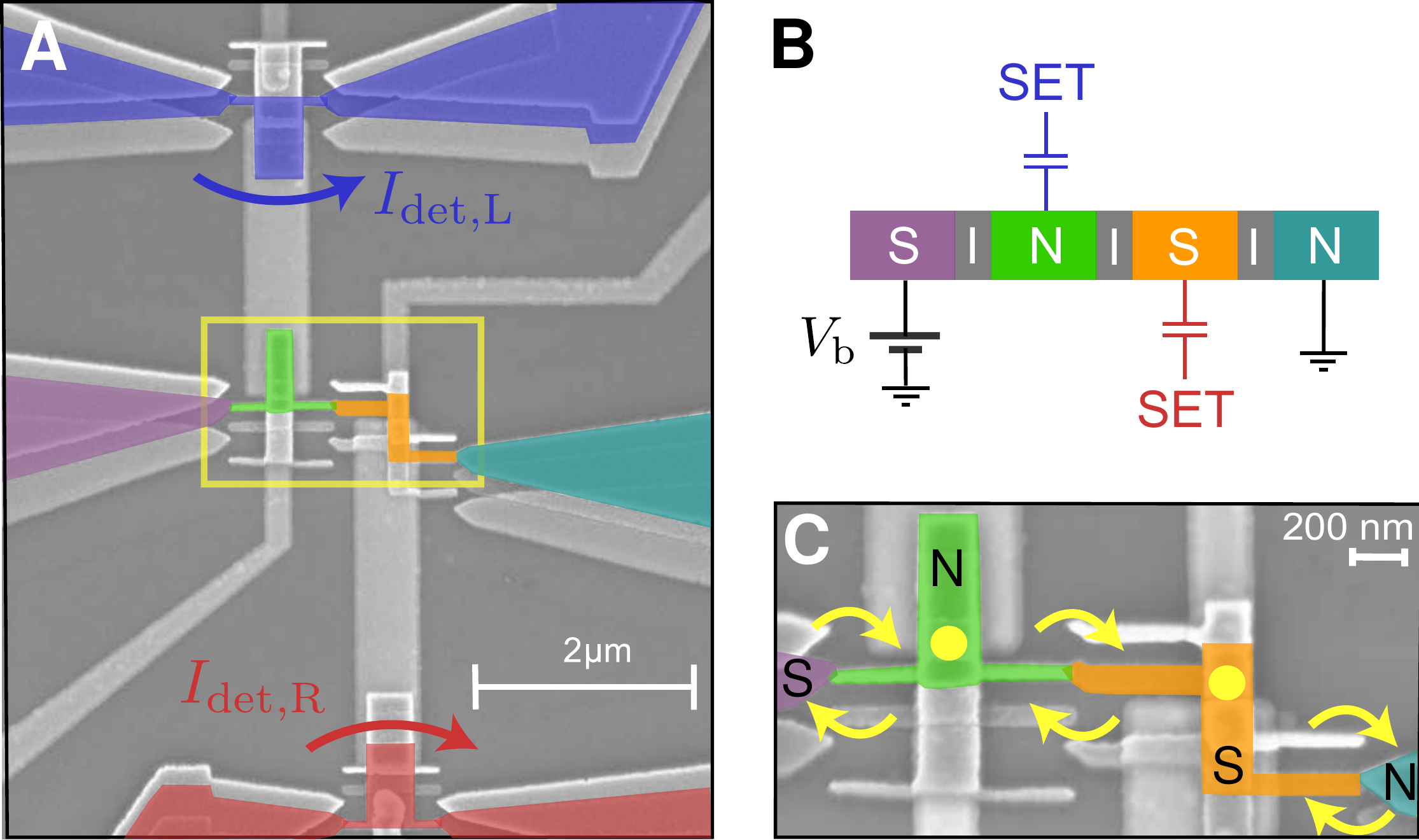} 
\caption{Experimental realization of a double dot.
(A)~Scanning electron micrograph of the sample with false color identifying its different components. The device consists of two leads, left (purple) and right (turquoise), two islands, left (green) and right (orange), and two single-electron transistor (SET) detectors, left (blue) and right (red).
(B)~Sketch of the circuit elements of the sample (N, normal metal; S, superconductor; I, insulator) with colors corresponding to the ones in panel (A).
An external DC voltage $V_{\rm b}$ controls the net current through the double-dot.
(C)~Zoomed view of the yellow rectangular region in A. Electrons (yellow circles) can tunnel between the leads and the islands in the directions indicated by the arrows.
Details on fabrication techniques and measurement setup can be found in Appendix~\ref{a1} and \ref{a2}, respectively. \looseness-1
}
\label{fig:1}
\end{figure}

In this paper, we experimentally study fluctuation-induced negative changes of the  entropy production
in a hybrid normal metal$-$superconductor double dot in the strong
Coulomb blockade regime (see Fig.~\ref{fig:1}). We demonstrate that the average magnitude
of such changes lies above the universal negative lower bound $-k_{\rm B}$ in agreement with the theoretical
prediction~\cite{PhysRevX.7.011019}. This remarkable result applies generally to any system
in non-equilibrium steady state. Furthermore, we derive and test in the experiment the upper bound on the average amount
of energy, which the system extracts from the environment during a negative entropy production fluctuation.
Interestingly, this bound is not universal and may significantly exceed $k_{\rm B}T$, where $T$ is the temperature.
We also perform more detailed comparisons between the theory and the experiment on the level of
statistical distributions of the minima of the entropy production.

\begin{figure}[ht!]
 \includegraphics[width= 0.7 \linewidth]{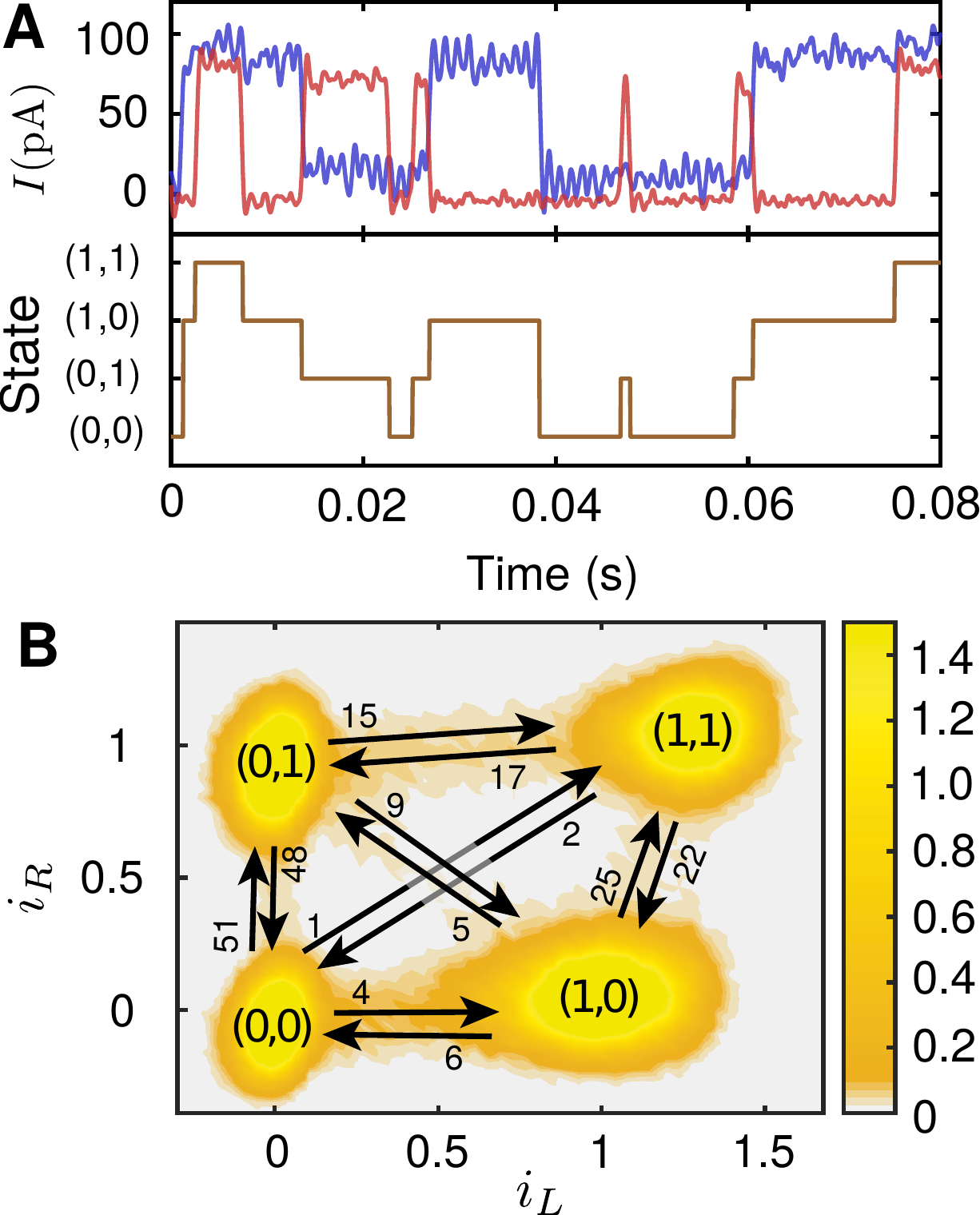} 
\caption{Experimental measurements on  
a double dot. (A)~Top: time trace of left ($I_{\text{det},L}$, blue) and right ($I_{\text{det},R}$, red) detector currents for $V_{\rm b}=90\;\mu\rm{V}$. Bottom: corresponding time trace for the charge state ($n_L,n_R$) of the double dot, where $n_L$($n_R$) = 0 or 1 implies the absence or presence of one extra electron in the left (right) island, respectively (see Appendix~\ref{ana} for details).
 (B)~Probability density of the normalized detector currents ($i_L,i_R$)~\cite{nor_cur} obtained from a $15\,\rm s$ time trace for $V_{\rm b}=90\;\mu\rm{V}$, showing four densely populated charge states~($n_L,n_R$).
The black arrows represent the possible transitions between the states with the numbers indicating the number of jumps per second averaged over $7.5$~h.}
\label{fig:1curr}
\end{figure}
Theoretical results relevant for our experiment are based on extreme-value statistics~\cite{gumbel2012statistics}.
Understanding extreme-value statistics of stochastic processes has attracted considerable attention in several disciplines of science such as finance, climate physics and DNA replication~\cite{gumbel2012statistics,PhysRevE.78.041917,kotz2000extreme,mcneil2000estimation}. Key concepts are the maximum and minimum of a stochastic process $X(t)$ over a finite-time interval $[0,t]$, which are given by $X_{\rm max}(t) \equiv \max_{t'\in[0,t]} X(t')$ and $X_{\rm min}(t) \equiv \min_{t' \in [0,t]} X(t')$, respectively.
Note that $X_{\rm max}(t)$ and $X_{\rm min}(t)$ are, respectively, increasing and decreasing stochastic processes. Universal extreme-value distributions in stationary stochastic processes have been found in the context of random walks~\cite{1742-5468-2008-05-P05004,krug2010,krug2011,vivo2015,satyareview,benichou2016temporal,finch2018far} and stochastic thermodynamics~\cite{PhysRevX.7.011019,Chetrite2011,PhysRevLett.119.140604}.
Recent theory has investigated generic bounds for the probability that the minimum of entropy production falls below a certain value.
This result was related to statistics of the maximal number of steps that a hopping process can move against a thermodynamic bias~\cite{PhysRevX.7.011019,Chetrite2019martingale}.
An important experimental test bench for this physics are single-electron devices in which stochastic transfer of electrons in the presence of an electric bias can be measured \cite{PekolaRMP2013,koski644,OPS_PRL.109.180601,PhysRevB.94.241407}.
However, the statistics of entropy-production extrema has  
not been investigated in device physics and their implications to single-electron transport remain yet unknown.\looseness-1

We measure nonequilibrium charge-state fluctuations in a hybrid normal metal$-$superconductor double dot in the strong
Coulomb blockade regime subject to a time-independent bias voltage (Fig.~\ref{fig:1}). The device is highly resistive, and electron tunnelling rates are therefore low ($\sim100$~Hz) compared with the sampling rate $f_{\rm s} = 25$~kHz.
Two single-electron transistor (SET) detectors, each one coupled to each of the two dots, ensure a sufficient signal-to-noise ratio for a reliable detection of every single-electron tunnelling event, as has been demonstrated before~\cite{koski644,kung2012irreversibility,PekolaRMP2013}.
Counting charges in single-island devices
does not provide information on the direction of electron transport,
a key feature to measure entropy production.
Our double dot provides more information, enabling
the measurement of the direction of single-electron currents~\cite{Fujisawa1634,kung2012irreversibility} and thus of time traces of stochastic entropy production $S(t)$, as we show below~\cite{Seifert040602,Lebowitz1999}.
Using this data we study the extreme-value statistics of $S(t)$ and relate it to recent theoretical predictions~\cite{PhysRevX.7.011019,Chetrite2011,PhysRevLett.119.140604}. We furthermore discuss how the extreme-value statistics of $S(t)$ can be related to the extreme-value statistics of heat exchanged by the device with its environment under isothermal conditions.

This paper is organized as follows. In Sec.~\ref{s2} we describe how stochastic entropy production can be evaluated from steady-state charge fluctuations of a double dot and report on its experimental measurement. In Sec.~\ref{s3} we discuss experimental results on extreme-value statistics of stochastic entropy production and compare our results with theoretical predictions. In Sec.~\ref{s4} we extend our theory to describe extreme-value statistics of heat and environmental entropy changes and test our theory with experimental data. In Sec.~\ref{s5} we provide insights on how our theory can be extended to nonequilibrium systems that are driven out of the steady-state regime and relate this theory to our experimental results, and Sec.~\ref{s6} contains the discussion.
Finally, in the appendices we discuss the fabrication technique~\eqref{a1},
measurement setup~\eqref{a2},
physics of double dot and detector back-action~\eqref{Teff_cal},
data analysis~\eqref{ana}, and general bounds for heat extrema~\eqref{a5}.

\begin{figure*}[ht]
	\centering
	\includegraphics[width=0.75\linewidth]{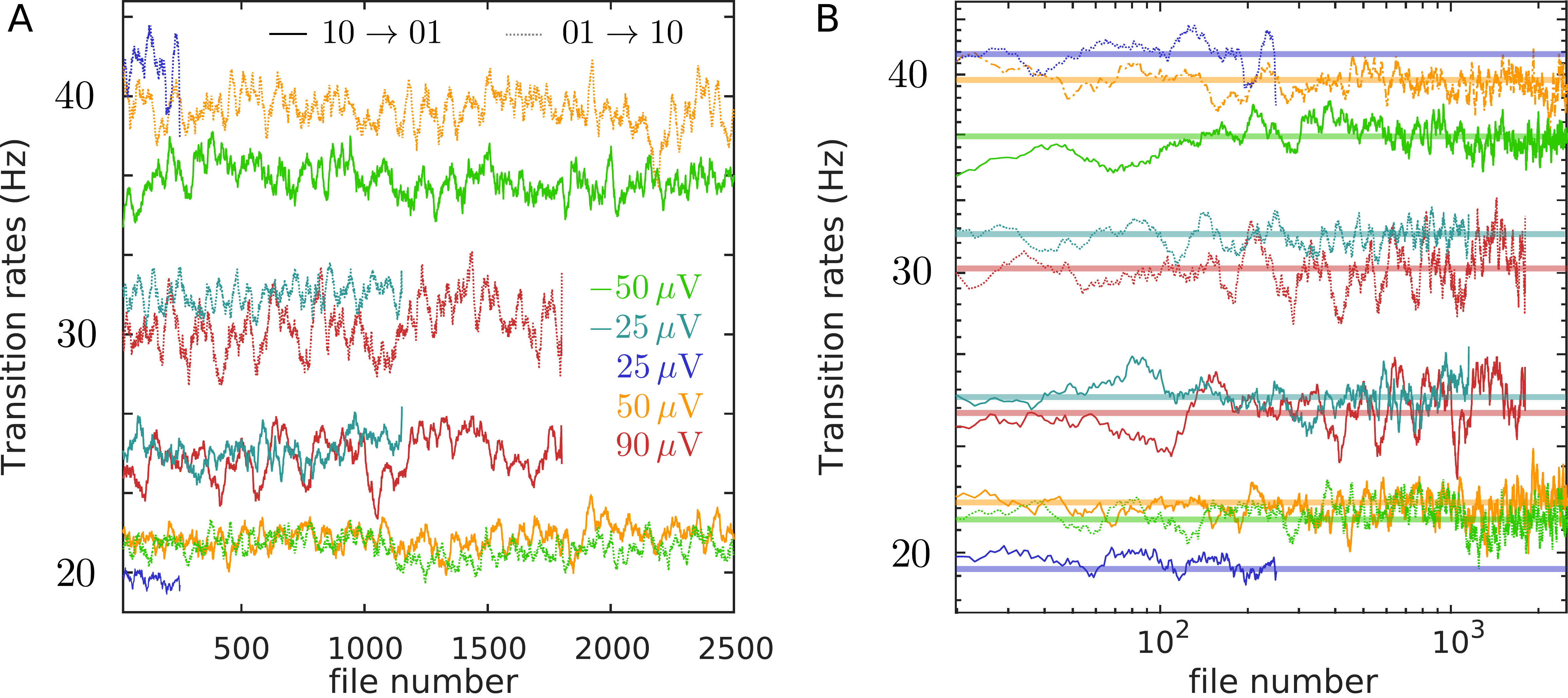}  
	\caption{ { Fluctuations in transition rates over time.}
		The  transition rates $\Gamma_{10, 01}$ (solid lines) and $\Gamma_{01,10}$ (dotted lines) as a 		function of  file numbers for different bias voltages (see legend in A).
		Each file is a $15$~s time evolution of system charge state. 
		The transition rates are evaluated using Eq.~\eqref{eq:Tr} (as described in Sec.~\ref{s2} and Appendix~\ref{ana})
		and the variables $N_{n\to n'}$, $P^{\rm st}(n)$ computed for the individual files with time duration $\tau = 15$~s. 
		The solid line in B is the average transition rate (shown in Table~\ref{tab:t1}) used for computing the entropy production using Eqs.~(\ref{eq:1}-\ref{eq:12}).
		\looseness-1
		\label{fig:SI6} }
\end{figure*}

\section{Experimental measurement of Stochastic entropy production}
 \label{s2}
To quantify extreme statistics at high resolution we use a custom-built electronic double dot. Our experimental setup
consists of two metallic islands tunnel-coupled to two leads and to each other, and capacitively coupled to two charge detectors (Fig.~\ref{fig:1}A).
An external DC bias voltage $V_{\rm b}$ is applied between
the two leads and brings the system into a nonequilibrium steady state (Fig.~\ref{fig:1}B). The system exhibits single-electron currents where electrons tunnel between leads and islands (Fig.~\ref{fig:1}C). In order to infer these fluctuating currents from the experimental data we describe the double dot as a four-state system $n=(n_L, n_R)$ with $n_{L,R}\in\{0,1\}$ as the left and right Coulomb-blockaded islands can be occupied by either zero or one extra electron, see Appendix~\ref{a2}.
The charge state $n_{L,R}$ of each island is detected by the SET coupled to the corresponding island (Fig.~\ref{fig:1curr}A) and thus each single experimental realization is characterized by stochastic trajectories of duration $t$ of the charge state $\{n(t')\}_{t'=0}^t$. From these Markovian trajectories (see Appendix~\ref{ana} for details) we
quantify the mesoscopic time-integrated currents $J_{m,m'}(t)$~(Fig.~\ref{fig:1curr}B). These currents are defined as the net number of transitions between states $m$ and $m'$ during a time interval $[0,t]$ in the trajectory $\{n(t')\}_{t'=0}^t$~\cite{PhysRevB.88.115134,1367-2630-15-12-125001}.

We analyze the nonequilibrium charge-transport fluctuations in the double dot using the framework of stochastic thermodynamics. Specifically,
 we measure the stochastic entropy production $S(t)$ associated with a given charge-state trajectory of the double dot $\{n(t')\}_{t'=0}^t$.
For stationary Markov jump processes~\cite{ftnt_dummies}, $S(t)$ is defined as a linear combination of the currents~\cite{Lebowitz1999,Seifert040602}
\begin{equation}
S(t) = \Delta S^{\rm sys}(t)+S^{\rm e}(t)\quad,
\label{eq:1}
\end{equation}
with the system entropy change~\cite{Seifert040602,gavrilov2017direct}
\begin{equation}
\Delta S^{\rm sys}(t) = \sum_{m<m'}\! \Delta S^{\rm sys}_{m,m'}\, J^{}_{m,m'}(t)\quad,
\label{eq:13}
\end{equation}
 and the entropy flow to the environment~\cite{prigogine1989entropy}
 \begin{equation}
 S^{\rm e}(t) = \sum_{m<m'}\! S^{\rm e}_{m,m'}\, J^{}_{m,m'}(t)\quad.
 \label{eq:1Senv}
 \end{equation}
The parameters $\Delta S^{\rm sys}_{m,m'} $ and $ S^{\rm e}_{m,m'}$ in Eqs.~(\ref{eq:1}-\ref{eq:1Senv}) are thermodynamic forces
\begin{equation}
\Delta S^{\rm sys}_{m,m'} = \log\, \left(\frac{P^{\rm st}_m}{P^{\rm st}_{m'}}\right)\;,\; S^{\rm e}_{m,m'}= \log\, \left(\frac{\Gamma_{m, m'}}{ \Gamma_{m', m}}\right)
\label{eq:12}
\end{equation}
defined as the change of mesoscopic system entropy and mesoscopic entropy flow to the environment during the transition $m\to m'$. Here we have defined the stationary probability to be in state $m$ as $P^{\rm st}_{m} = \langle t_m \rangle/t $, where $t_m$ is the occupation time in state $m$ and $t$ is the total duration of the trace. We also define the transition rates from states $m$ to $m'$ as $\Gamma _{m, m'}= \langle J_{m, m'}(t)\rangle/( P^{\rm st}_m t )$,
 where $\langle \,\cdot\, \rangle$ denotes here an average over many realizations.
Here and further we use $k_{\rm B}=1$ and the natural logarithm by $\log$.

The definition (\ref{eq:1}) implies that at
thermodynamic equilibrium $S(t)=0$
whereas in a nonequilibrium steady state
both $S(t)$ and $S^{\rm e} (t)$ increase with time {\em on average},
$\langle S(t)\rangle > 0$ and $\langle S ^{\rm e} (t)\rangle > 0$,
 in agreement with the second law of thermodynamics. If the environment consists of several thermal reservoirs and local detailed balance holds, the mesoscopic entropy flow to these reservoirs $S^{\rm e}(t)=-\sum_k Q_k(t)/T_k$, where $-Q_k(t)$ is the heat dissipated to
a thermal reservoir at temperature $T_k$, see Appendix~\ref{Teff_cal}.
Fluctuations of entropy production have also universal features.
 The most studied examples are fluctuation theorems, which imply that negative values of $S(t)$ occur exponentially less often than events with positive $S(t)$: the probability distribution of stochastic entropy production at a given time $t$ is asymmetric around zero, $P(S(t)=-s)=P(S(t)=s)\exp(-s)$~{\cite{seifert2012stochastic}}.
 As a consequence, the cumulative distribution of the stochastic entropy production obeys the inequality:
\begin{equation}
\mathsf{Pr}\left(S(t)\geq -s\right) \geq 1-\exp(-s),\quad s\geq 0, \label{eq:udo}
\end{equation}
where $\mathsf{Pr}\left(\cdot\right)$ denotes the probability of an event~\cite{seifert2012stochastic}.
 For Markovian systems, events of entropy reduction are associated with transitions $m'\to m$ against the direction of thermodynamic forces~\cite{PhysRevB.88.115134,1367-2630-15-12-125001} for which $ J_{m,m'}(t)$ decreases transiently, e.g. when
an electron travels in the direction opposite to the electric force.

\begin{figure}
\centering
 \includegraphics[width=0.76\linewidth]{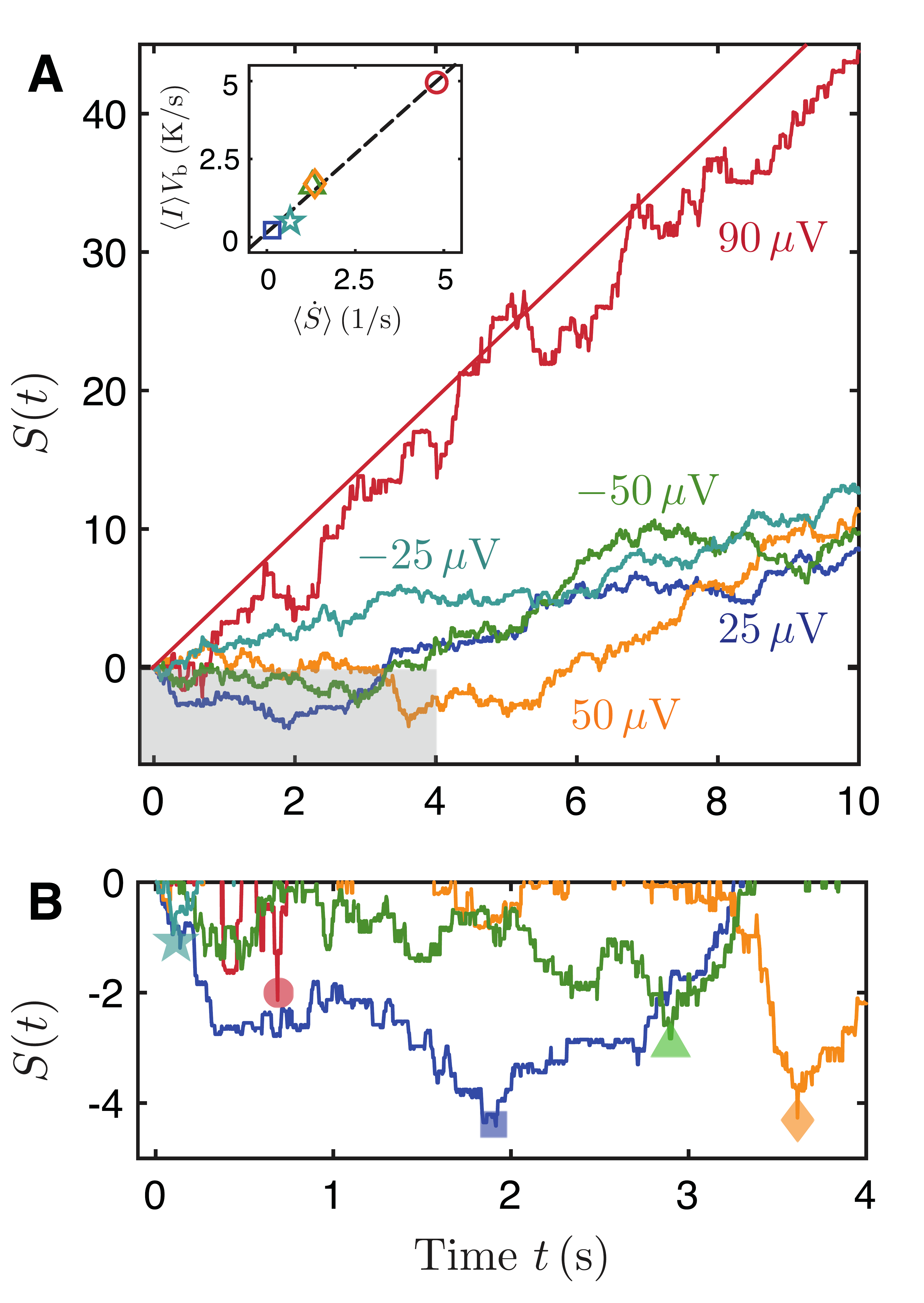}
\vspace{-0.3cm}
\caption{Experimental measurement of stochastic entropy production.
(A)~Sample traces of stochastic entropy production as a function of time, for different values of the bias voltage (see legend).
The straight red line is the average Joule heating for $90~\mu$V normalized by $T_{\rm eff}$.
Inset: Steady-state average Joule dissipation rate $\langle I\rangle V_{\rm b}$ as a function of the steady-state entropy production rate $\langle\dot{S}\rangle $ for different bias voltages. The dashed line is a linear fit with slope $T_{\rm eff} = (1.01\pm 0.16)\,\rm K$, y-intercept $y_0=(0.14\pm 0.40)\,\text{K}/s$ and $R^2=0.990$, see Eq.~\eqref{eq:2}.
(B)~Zoomed view of the shaded region in (A) with finite-time minima $S_{\rm min}(t)$ of each trace represented with symbols.
\label{fig:2} }
\end{figure}

\begin{figure*}[ht]
\centering
 \includegraphics[width=0.9\linewidth]{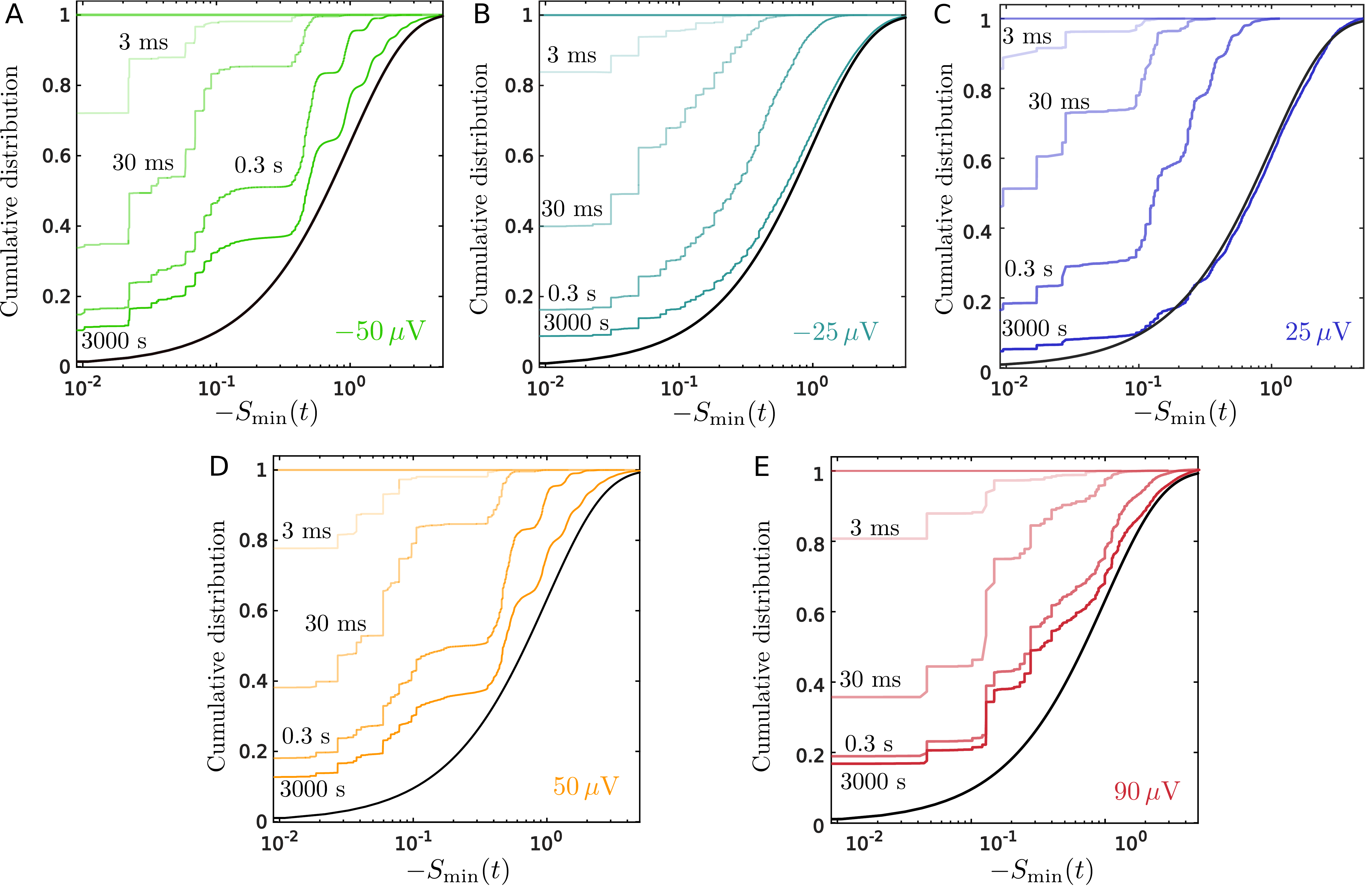}
\caption{ Experimental cumulative distribution of the finite-time entropy production minima in the double dot with bias voltage $V_{\rm b}=-50\;\mu\rm{V}$~(A),
$-25\;\mu\rm{V}$~(B), $25\;\mu\rm{V}$~(C), $50\;\mu\rm{V}$~(D), and $90\;\mu\rm{V}$~(E) for different values of the observation time (see legend). A horizontal line is set to one to guide the eye and corresponds to the cumulative distribution at $t=0$. The black curve is the theoretical bound given by the right-hand side in Eq.~\eqref{eq:3}.
\looseness-1
\label{fig:3} }
\end{figure*}

In our experiment, we first measure $P^{\rm st}_{m}$ and $\Gamma_{m, m'}$ from the counting statistics of a large ensemble of charge-state traces.
Note that the empirical estimates of $P^{\rm st}_{m}$ and $\Gamma_{m, m'}$ are affected by different sources of noise. First, they may vary with time due to finite-time statistics. Another effect comes from feedback control which ensures only approximatively nonequilibrium stationary conditions as, for some experiments, we observe residual drift effects. These two effects may affect the counting statistics and therefore the estimation of both the stationary probabilities and the transition rates (see Fig.~\ref{fig:SI6}A).
We use time-averaged $P^{\rm st}_{m}$ and $\Gamma_{m, m'}$ (Fig.~\ref{fig:SI6}B) and fluctuating charge-state trajectories to quantify both $ S (t) $ and $ S^{\rm e} (t)$ using Eqs.~(\ref{eq:1}-\ref{eq:12}).
We remark that $ S (t) $ is a {\em functional} that associates to each charge-state trajectory $\{n(t')\}_{t'=0}^t$ another stochastic trajectory $\{S(t')\}_{t'=0}^t$, with the estimated steady-state values of $P^{\rm st}_{m}$ and $\Gamma_{m, m'}$ being parameters of the functional.

We then plot traces $\{S(t')\}_{t'=0}^t$ of stochastic entropy production from the experimental data
of the double dot for different values of the bias voltage~$V_{\rm b}$ ranging from $-50~\mu V$ to $90~\mu V$ (Fig.~\ref{fig:2}A). Trajectories $\{S(t')\}_{t'=0}^t$
exhibit transiently negative values but increase with time on average, as expected from the second law. The average rate of entropy production is linearly proportional to the Joule dissipated power~\cite{Andrieux2008JSM,PhysRevLett.98.150601} in the double dot,
\begin{equation}\label{eq:2}
 \langle I \rangle V_{\rm b} =T_{\rm eff} \langle \dot{S} \rangle \quad,
\end{equation}
 see inset in Fig.~\ref{fig:2}A.
In Eq.~\eqref{eq:2} the average electric current between the two islands is defined as $\langle I\rangle = e\,[P^{\rm st}_{(0,1)}\Gamma_{( 0,1),(1,0)} - P^{\rm st}_{(1,0)}\Gamma_{( 1,0) , (0,1) }]$ with $e$ the elementary charge.
The parameter $T_{\rm eff}\simeq 1\,\rm K$ is an effective temperature that characterizes the nonequilibrium nature of the environment. It is one order of magnitude larger than the base temperature ($T=50$~mK) and the electronic temperature of the superconducting and normal-metal components $T_{\rm el}\approx 170$~mK, see Appendix~\ref{Teff_cal}.
The main contribution to the difference between $T_{\rm eff}$ and $T_{\rm el}$
is given by backaction of the detectors, strongly coupled to the sample and operated away from equilibrium, see Appendix~\ref{Teff_cal}.
Note that in earlier experiments where the detector backaction was minimized, temperatures $T_{\rm eff} \sim T_{\rm el} < 150$~mK have been reported~\cite{OPS_PRL.109.180601,koski2014experimental} with the same type of detectors having weaker dot-detector coupling.

\section{Extreme values of stochastic entropy production}
\label{s3}

\begin{figure}[t]
\centering
 \includegraphics[width=70mm]{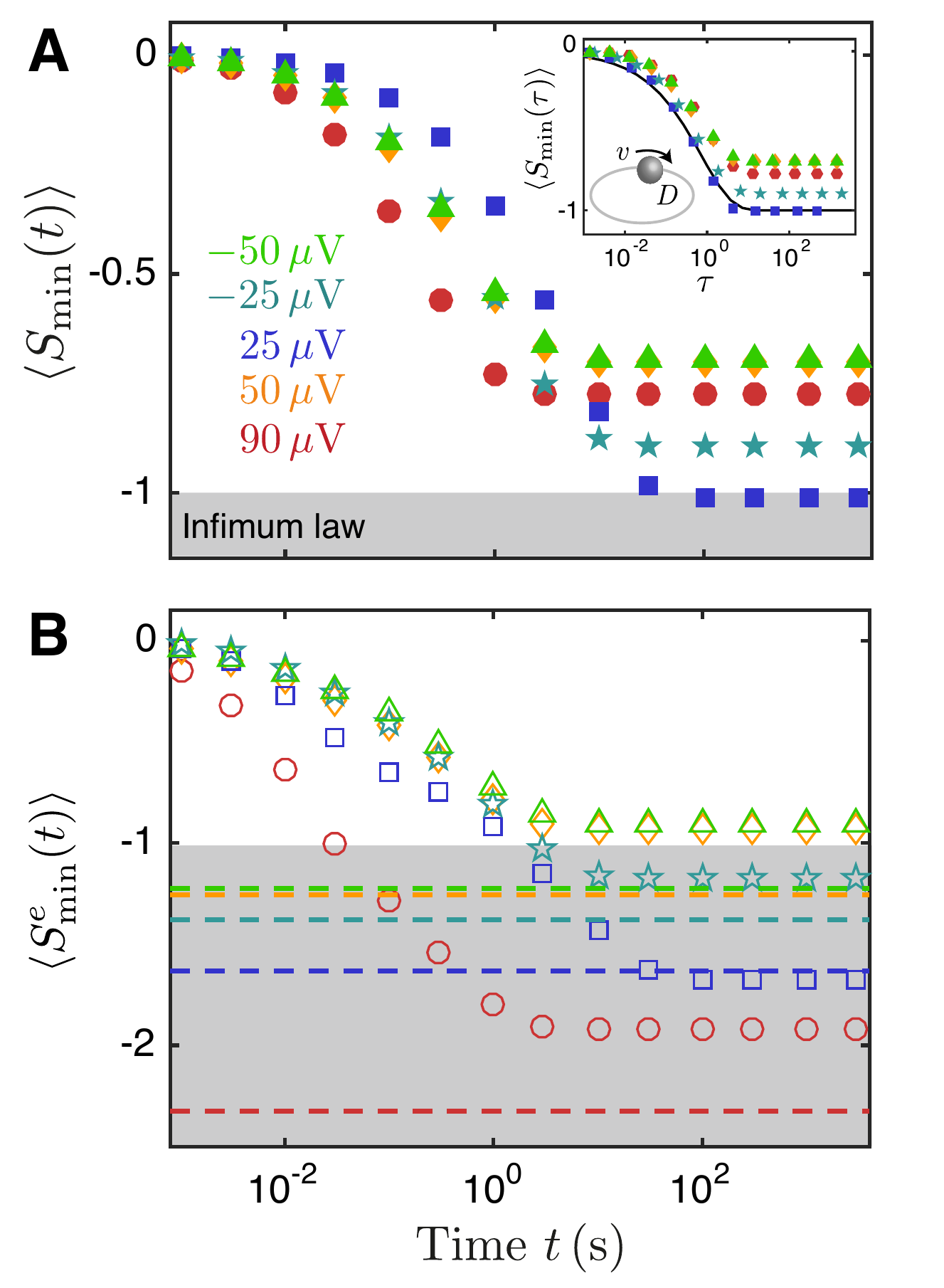}
\caption{ Finite-time average minimum of stochastic entropy production $\langle S_{\rm min}(t)\rangle$~(A) and entropy outflow $\langle S^{\rm e}_{\rm min}(t)\rangle$~(B) in the double dot
 as a function of time, for different values of the bias voltage (shown in different colors). The grey area indicates the events where entropy extrema are below $-1$ and the
 dashed lines in (B) are given by the right-hand side of Eq.~\eqref{eq:5}. The inset in (A) shows the value of $\langle S_{\rm min}(\tau)\rangle$ as a function of $\tau = t \langle \dot{S}\rangle$, where the black curve is the average entropy-production minimum for a driven colloidal particle in a ring with equal drift and diffusion coefficients $v=D=1$, given by $-\text{erf}(\sqrt{\tau}/2) +(\tau/2)\text{erfc}(\sqrt{\tau}/2)- \sqrt{\tau/\pi}\exp(-\tau/4)$.
 }
\label{fig:4}
\end{figure}

From the experimental traces of stochastic entropy production, we measure
the minimum value of stochastic entropy production over a finite time $t$, $S_{\rm min}(t) = \min_{t'\in[0,t]} S(t')$, which is a negative random variable $S_{\rm min}(t)\leq 0$ , since $S(0)=0$ (Fig.~\ref{fig:2}B).
Next, we collect statistics of such negative extreme values over many traces of fixed duration $t$, and plot the cumulative distribution function of $S_{\rm min}(t)$ for different bias voltages, ranging from $-50~\mu V$ to $90~\mu V$. 
Remarkably, the experimental cumulative distributions of the finite-time minima of stochastic entropy production (Fig.~\ref{fig:3}A-E), can be bounded, for all the experimental conditions by  
a universal exponential distribution
\begin{equation}
\mathsf{Pr}\left(S_{\rm min}(t)\geq -s\right) \geq 1-\exp(-s),\quad s\geq 0, \label{eq:3}
\end{equation}
for all values of $t$, in agreement with recent theory for nonequilibrium steady states~\cite{PhysRevX.7.011019,Chetrite2011}. Thus, the tail of the distribution of entropy-production minima is suppressed, stronger than exponentially, in the thermodynamically forbidden region $S_{\rm min}(t)<0$. Moreover, the experimental average minimum of stochastic entropy production (Fig.~\ref{fig:4}A) obeys the so-called infimum law~\cite{PhysRevX.7.011019}\looseness-1
\begin{equation}
\big\langle S_{\rm min}(t)\big\rangle \geq -1\quad, \label{eq:4}
\end{equation}
as follows from Eq.~\eqref{eq:3}.
Note that, for $V_{\rm b}=25\;\mu\rm{V}$, the empirical long-time average minimum
 $\langle S_{\rm min}(\infty)\rangle \simeq (-1.01\pm 0.02)$ which is in agreement with Eq.~\eqref{eq:4}.
For this case, the bound is tight because entropy production jumps given by Eq.~\eqref{eq:12} are $\ll 1$, see Appendix~\ref{Teff_cal} and the traces of $S(t)$ can be approximated by
those of a continuous stochastic process, for which $\langle S_{\rm min}(\infty)\rangle = -1$~\cite{PhysRevX.7.011019,PhysRevLett.119.140604}.
Interestingly, the finite-time average minimum can be lower bounded, for all bias voltages, by a master curve when rescaling time by the entropy production rate $\tau=t\langle\dot{S}\rangle$ (Fig.~\ref{fig:4}A inset).
Such master curve is given by the average minimum of the position of a 1D drift-diffusion process with equal drift and diffusion coefficients $v=D=1$~\cite{PhysRevX.7.011019,marzuoli2015extreme}.

\section{Extreme values of environmental entropy changes and heat}
\label{s4}

We now demonstrate that Eq.~\eqref{eq:4} implies also a bound for the mean of
the minimum mesoscopic entropy flow to the environment, and test the
implications of this theoretical result with experimental data.
First, from Eqs.~\eqref{eq:1} and~\eqref{eq:4}, we derive in Appendix~\ref{a5} the following bound for
the average minimum of the entropy flow $S^{\rm e}_{\rm min}(t)=\min_{t' \in [0,t]} S^{\rm e}(t')$
\begin{equation}\label{eq:5}
\langle S^{\rm e}_{\rm min}(t) \rangle \geq - 1 - \sum_{n} P^{\rm st}_n \log (P^{\rm st}_n/P^{\rm st}_{\rm min}) \quad.
\end{equation}
Here we have defined $P^{\rm st}_{\rm min} = \min_{n'} P^{\rm st}_{n'}$.
Since
 $P^{\rm st}_{\rm min}\leq P^{\rm st}_n$ for all states $n$, the second term in Eq.~\eqref{eq:5} is negative and therefore the average of the minimum value of $S^{\rm e}(t)$ can be smaller than $-1$.
 Our experimental results are in agreement with the bound~\eqref{eq:5}, for all tested values of the bias voltage $V_{\rm b}$ (see Fig.~\ref{fig:4}B).
Equation~\eqref{eq:5} implies that the average minimum of the entropy flow can be below~$-1$ for steady states with heterogeneous probability distributions, as is the case for $V_{\rm b}=\pm 25\;\mu\rm{V}$ and $V_{\rm b}=90\;\mu\rm{V}$.

 From Eq.~\eqref{eq:5} and using $S^{\rm e}(t)=-Q(t)/T$, we predict that the
 maximum value of the heat that an isothermal mesoscopic system can absorb from its environment in a time interval $[0,t]$ cannot exceed on average
\begin{equation}\label{eq:6}
 \langle Q_{\rm max}(t)\rangle \leq T \left[ 1 + \sum_n P^{\rm st}_n \log (P^{\rm st}_n/ P^{\rm st}_{\rm min})\right] \quad.
\end{equation}
Interestingly the bound (\ref{eq:6}) holds for all mesoscopic systems in a nonequilibrium steady state, regardless of the system size and complexity.
This fundamental limit, of the order of $T$, is comparable to average work extracted by mesoscopic information engine (e.g. Szilard) in a single cycle~\cite{toyabe2010experimental,roldan2014universal,koski2014experimental,gavrilov2017direct}. For systems in contact with nonequilibrium environments where local detailed balance is approximatively satisfied at an effective temperature, $S^{\rm e}(t)\simeq -Q(t)/T_{\rm eff}$,  
 one can estimate the average maximum heat replacing $T $ by $T_{\rm eff}$ in Eq.~\eqref{eq:6}.

 \section{Statistics of Extreme entropy reductions out of steady state}
 \label{s5}

 In our experiment, the transition rates between different charge states fluctuate over time as a result of the shot noise in the detectors that are strongly coupled to the double dot (Fig.~\ref{fig:SI6}). The results shown in Figs.~\ref{fig:3} and~\ref{fig:4} were obtained using time-averaged values of the rates in Eqs.~(\ref{eq:1}-\ref{eq:12}). The agreement between the theoretical predictions and the experimental results thus implies that the experiment realizes in very good approximation a non-equilibrium steady state with rates given by their time-average rates obtained from long charge-state trajectories. If the nonequilibrium conditions are far from stationary, extreme statistics of stochastic entropy production may not obey the bound~\eqref{eq:4} as we show below.

First we observe that the condition (\ref{eq:3}) follows from the fact that for a Markovian and stationary process $n(t)$ the exponent
$\exp(-S(t))$ is a martingale process. For nonstationary processes $\exp(-S(t))$ is no longer a martingale process, and therefore (\ref{eq:3}) does not hold in general.
Next, by definition $S_{\rm min}(t)\leq S(t)$, and hence
\begin{equation}
\mathsf{Pr} \left(S(t)\geq -s\right) \geq \mathsf{Pr}\left(S_{\rm min}(t)\geq -s\right).
\label{eq:dummies}
\end{equation}
Therefore in a steady state the general bound for the entropy distribution $\text{Pr}(S(t)=s)$~\eqref{eq:udo} follows from
the condition~(\ref{eq:3}) for the distribution of the entropy infimum $\text{Pr}(S_{\rm min}(t)=s)$, but {\em not vice versa}.

\begin{figure}[t]
\centering
 \includegraphics[width=74mm]{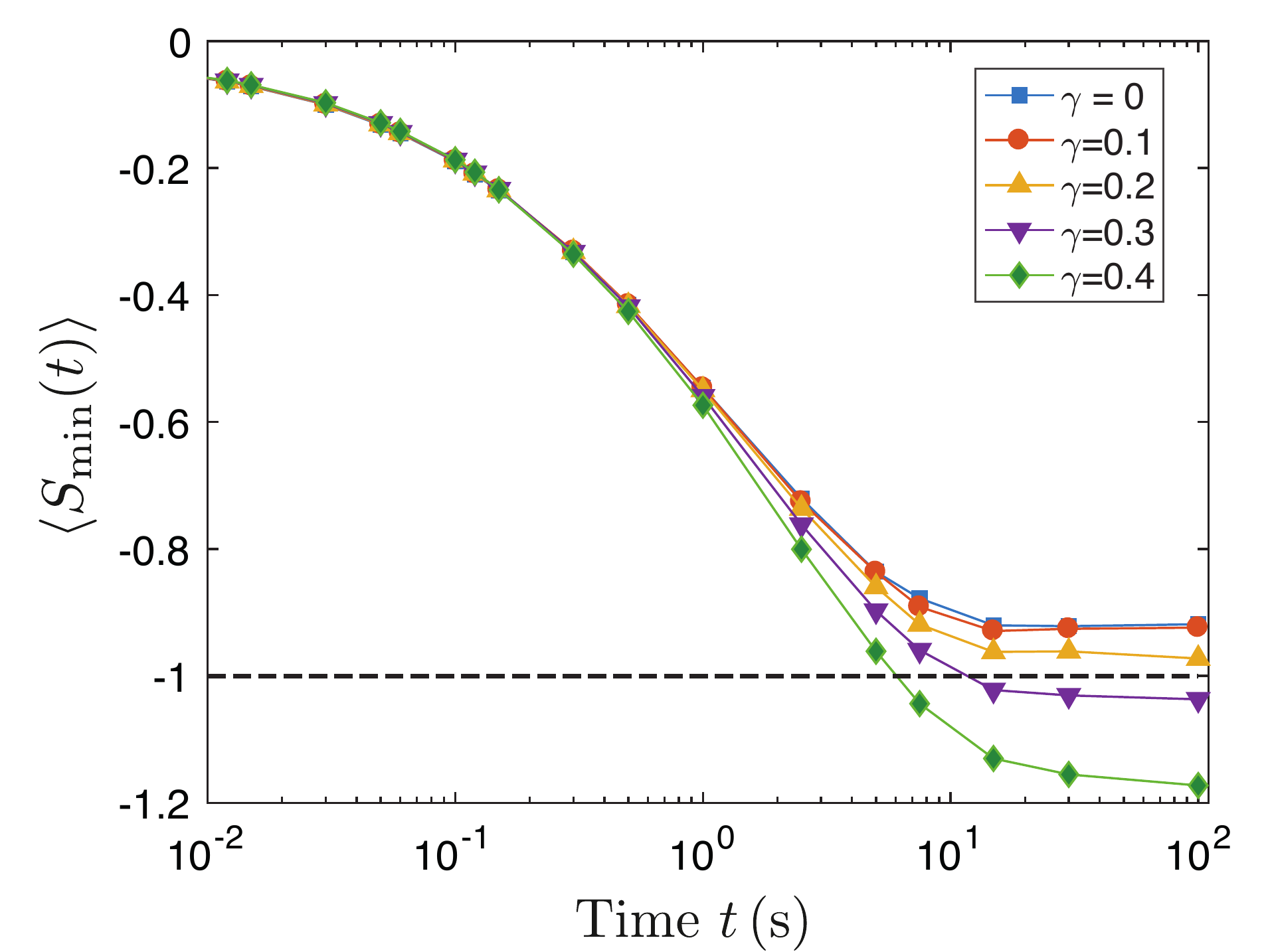}
\caption{ Finite-time average minimum of stochastic entropy production $\langle S_{\rm min}(t)\rangle$ as a function of time (symbols) obtained from numerical simulations of a double dot driven out of equilibrium and with time-dependent rate $\Gamma_{(0,1),(1,0)}(t) =\Gamma_0[1+\gamma\theta(N(t))]$. Here $\Gamma_0 = 27.1$~Hz, $N(t)$ is the total number of jumps between any two states up to time $t$ and $\theta(N)$ denotes the parity of $N$, i.e. $\theta(N)=1$ ($\theta(N)=-1$) for $N$ even (odd). The parameter $\gamma>0$ is fixed in the simulations and controls the amplitude of the variation of the time-dependent rate (see legend). The values of the time-independent rates (in Hz) are $\Gamma_{(0,0),(0,1)} = 101.8,\; \Gamma_{(0,1),(0,0)} = 2,\; \Gamma_{(0,0),(1,1)} = 1.6, \; \Gamma_{(1,1),(0,0)} = 74.1,\; \Gamma_{(0,0),(1,0)} = 29.3, \; \Gamma_{(1,0),(0,0)} = 31 ,\; \Gamma_{(0,1),(1,1)} =22.5,\; \Gamma_{(1,1),(0,1)} = 44, ,\; \Gamma_{(1,0),(1,1)} =73.9,\; \Gamma_{(1,1),(1,0)} = 101.4,\; \Gamma_{(0,1),(1,0)}= 28$ and the total number of simulations is $5\times 10^4$ for all cases.}
\label{fig:5}
\end{figure}

To analyze the validity of the infimum law~\eqref{eq:4} in non-stationary conditions, we perform numerical simulations of a double dot with all the transition rates equal to those experimentally measured for $V_{\rm b}=25\,\mu \rm V$ except the rate $\Gamma_{(0,1),(1,0)}(t) $ that is changed over time. For simplicity, we choose the time-dependent protocol $\Gamma_{(0,1),(1,0)}(t) = \Gamma_0 (1+\gamma \theta(N(t)) ] $, with $\Gamma_0$ given by the time-averaged value of the rate $\Gamma_{(0,1),(1,0)}$ measured in the experiment, $\gamma\leq 1$ a factor that controls the amplitude of the driving, and $\theta(N(t))$ given by the parity ($\pm 1$) of the total number of jumps $N(t)$ that occur between any two states up to time $t$. Therefore, we switch after each jump the value of the rate $(0,1)\to(1,0)$ between the values $ \Gamma_0 (1\pm \gamma)$ and $ \Gamma_0 (1\mp \gamma)$.
We then calculate $S(t)$ associated to each trajectory of the system using in Eqs.~(\ref{eq:1}-\ref{eq:12}) the actual stationary distribution of the system and the values of the rates, using the time-averaged value $\Gamma_0 = \langle \Gamma_{(0,1),(1,0)}(t)\rangle $ to calculate the entropy production associated with the jumps $(0,1)\to (1,0)$ and $(1,0)\to (0,1)$ [see Eq.~\eqref{eq:12}].
 Figure~\ref{fig:5} shows that, when the amplitude of the driving increases ($\gamma\geq 0.3$), this procedure can yield values of the average minimum of stochastic entropy production below $-1$. This result suggests that one can use measurements of extreme reductions of stochastic entropy production (evaluated using time-averaged rates) to assess the quality of a non-equilibrium steady state, i.e. to quantify whether the underlying dynamics is stationary in good approximation.

 \section{Discussion}
 \label{s6}
We now discuss how the extreme statistics of entropy-production can be used to
characterize electronic devices. Consider, for example, a single-photon detector with a photo-current flowing in the direction opposite to the bias.
An absorbed photon would generate a negative current pulse $I(t)<0$ in it.
The same current pulse may be caused by current fluctuations, which would result in a dark count.
The corresponding entropy production is negative,
$-S_0=\int_0^{\infty} \text{d}t' I(t')V_{\rm b}/T_{\rm el} = -q_0 V_{\rm b}/T_{\rm el}$,
where $q_0$ is the absolute value of the total transferred charge.
The dark count occurs if the entropy production minimum crosses the value $-S_0$ during the pulse.
Hence, according to Eq.~\eqref{eq:3}, the dark count probability is limited by ${\rm Pr}(S_{\min}(t)<-S_0)\leq \exp(-q_0 V_b/T_{\rm el})$.
This bound is restrictive for extreme
fluctuations such that $\exp(-q_0 V_b/T_{\rm el})\ll 1$. This result complements the usual analysis of detector sensitivity
that only accounts for weak (Gaussian) current fluctuations and ignores extreme-value statistics.

Our experiment reveals that the probability for extreme reductions of stochastic entropy production in an electronic double dot is bounded in terms of an exponential distribution with mean equal to minus the Boltzmann constant,
for all observed bias voltages, as predicted by recent theory.
Interestingly, the bound~\eqref{eq:3} for extreme entropy reductions becomes tight in the linear response regime.
Our results demonstrate that, although the transition rates fluctuate in time, the experimental setup realizes to very good approximation a Markovian nonequilibrium stationary state.

Furthermore, we have shown with theory and experiment that the average extreme reduction of the entropy flow from the environment to a mesoscopic system is bounded in terms of a system-dependent quantity that depends on the heterogeneity in the stationary distribution.
 It would be interesting to explore the relevance of extreme heat statistics
in periodically-driven systems
with feedback control such as single-electron information engines
working close to the Landauer limit~\cite{PhysRevLett.115.260602}.
One could also extend this theory to quantum coherent systems, like quantum heat engines, providing bounds for the extreme heat absorption and work extraction~\footnote{For purely quantum systems there is a whole field of quantum thermodynamics considering, in particular, quantum effects on the efficiency of heat engines~\cite{ RevModPhys.81.1665,Campisi2011,Svensson2013NJP,Kosloff2014,Alicki1979,Scully2003,Quan2007,Campisi2014,Uzdin2015,Marchegiani2016,Hofer2016_PRB93,Campisi2016}.}.

 We acknowledge
 the provision of facilities by Aalto University at OtaNano Micronova Nanofabrication Centre
and the computational resources provided by the Aalto Science-IT project.
We thank Matthias Gramich and Libin Wang for technical assistance.
We acknowledge fruitful discussions with
Simone Pigolotti, Alexandre Guillet, Andre C. Barato, Vladimir E. Kravtsov,
Samu Suomela, Christian Flindt and Keiji Saito.
This work is partially supported
by Academy of Finland, Project Nos. 284594, 272218, and 275167
(S.~S., D.~S.~G., V.~F.~M., J.~T.~P., and J.~P.~P.),
 by European Research Council~(ERC) under the European Union's Horizon 2020 research
and innovation programme under grant agreement No. 742559 (SQH),
 by the Russian Foundation for Basic Research, German Research Foundation (DFG) Grant No. KH~425/1-1, and the Russian Science Foundation, Grant No. 17-12-01383 (I.~M.~K.). Correspondence and requests for materials should be addressed to S.~S.~(email: \url{sshilpi916@gmail.com}).
 \appendix

\section*{Appendix}
\section{Sample fabrication} \label{a1}

\begin{figure*}
 \includegraphics[width= 0.92\textwidth]{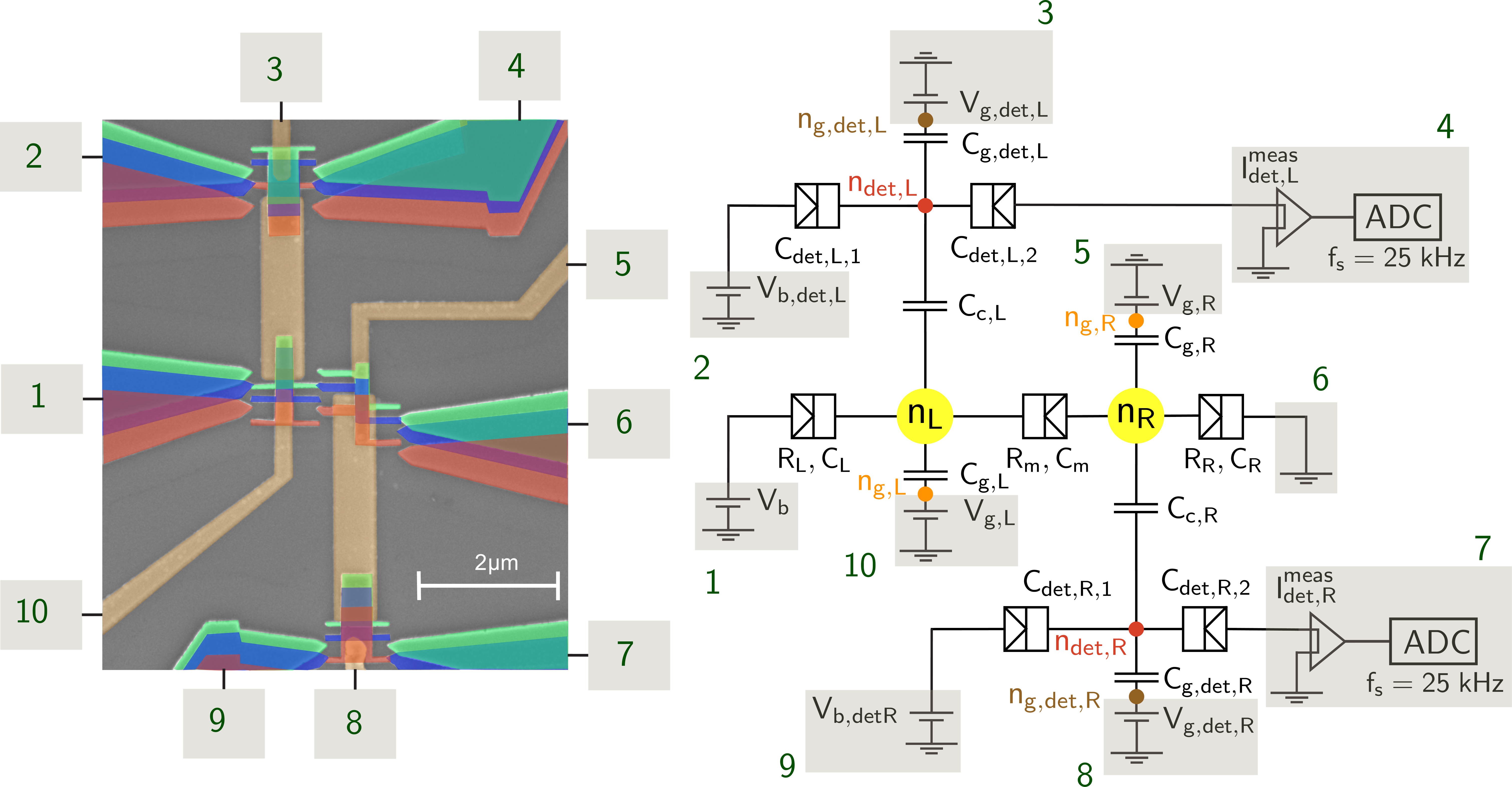}
\caption{\textbf{Sample and measurement scheme}. Left:~Sample micrograph with false color identifying the different stages of fabrication process.
Right:~Schematic of the electronic circuit corresponding to the sample together with the measurement setup.
The orange colored structures are ground plane, mainly $30$~nm of Au, evaporated before the $50$~nm of $ \rm Al_2 O_3$ dielectric layer. The orange rectangles are capacitive couplers between the double dot islands and the detectors islands, with capacitances $C_{\rm c,L}$ and $C_{\rm c,R}$. The remaining orange structures are gates to the double dot and detectors islands, each connected to a DC voltage source, $V_{\rm g,L}$, $V_{\rm g,R}$, $V_{\rm g, {det},L}$ and $V_{\rm g,{det},R}$
via the capacitances, $C_{\rm g,L}$, $C_{\rm g,R}$, $C_{\rm g, {det},L}$ and $C_{\rm g, {det},R}$, respectively.
The blue colored structures are Al evaporated as a first layer on the top of $ \rm Al_2 O_3$.
This Al is oxidized and covered by Cu (red) to make detectors' junctions, with capacitances $C_{\rm det,L,1}$, $C_{\rm det,L,2}$, $C_{ \rm det,R,1}$ and $C_{\rm det,R,2}$.
The Cu (green) is evaporated last to make high resistance double dot junctions, with capacitances $C_{ L}$, $C_{ m}$, $C_{ R }$ and resistances $R_{ L}$, $R_{ m}$, $R_{ R }$, corresponding to left, middle and right junction, respectively.
During the measurement, DC voltages, $V_{\rm {b}}$, $V_{\rm {b,det},L}$ and $V_{\rm {b,det},R}$ are applied across the double dot and detectors, through their left lead.
The currents $I_{\rm {det},L} ^{\rm meas}$ and $I_{\rm {det},R} ^{\rm meas}$ through detectors are recorded with sampling frequency $f_{\rm s} = 25$~kHz, by connecting an amplifier (triangles) and analog to digital converter (ADC) to the right lead of each of the detectors.
The circles represent the charge states of the dots ($n_{L}$ and $n_{R}$, yellow), the gates
($n_{g,L}$ and $n_{g,R}$, orange), the detector dots ($n_{\text{det},L}$ and $n_{\text{det},R}$, red) and the gates to the detectors ($n_{g,\text{det}, L}$ and $n_{g,\text{det},R}$, brown).
}
\label{fig:SI1}
\end{figure*}

{The experimental sample (see Fig.~\ref{fig:SI1}) consists of a double dot structure (left normal metal (N) island and right superconducting (S) island) consisting of three NIS junctions and of two single-electron transistors (SETs) used as detectors, fabricated following the Fulton-Dolan method~\cite{Dol77}.}
The fabrication process
{described below} consists of two electron beam lithography (EBL) steps, each followed by deposition of thin metal films by shadow evaporation~\cite{Wei01}.

 The first lithography
 step is needed for ground plane deposition (orange structures in Fig.~\ref{fig:SI1}). This is done as follows:
A polymer resist (approximately $300$~nm thick layer of positive e-beam resist ALLRESIST AR-P 6200) is prepared on top of the wafer by spin-coating a silicon substrate covered by $300$~nm thick layer of thermally-grown silicon oxide.
Then the wafer is exposed to $100$~kV electron beam for defining the gate electrodes and a continuous ground plane electrode to facilitate on-chip filtering of spurious microwave photons~\cite{PhysRevLett.105.026803}. A low beam current~($1$~nA) is used for small structures ($<6 \, \mu$m) that will be located close to the junctions,
and high current~($200$~nA) is used for large structures~(few $100 \, \mu$m) that form the pads for bonding different leads and gates.
After EBL, the exposed wafer is developed using developer AR 600-546, followed by isopropyl alcohol (IPA) rinse and N$_2$ dry.
The structures are metallized by evaporating $2$~nm of Ti, $30$~nm of Au and then $2$~nm of Ti.
The bottom Ti helps the Au to stick to the SiO$_2$. Then atomic layer deposition
technique is used to grow around $50$~nm thick $\text{Al}_2\text{O}_3$ dielectric layer on the wafer to isolate the ground plane from the bias leads and tunnel junction structures.

A second lithography step is applied for the fabrication of tunnel junctions using multi-angle shadow evaporation through a suspended mask.
For this step, a Ge-based mask is used~\cite{PekolaRMP2013}. The mask consists of three layers:
the topmost layer is approximately $50$~nm Polymethyl methacrylate (PMMA) (molecular weight $2.2$~million $1.8$\% in anisole),
the middle layer is Ge ($22$~nm) and
the bottom layer is $400$~nm methylmethacrylate (MMA) ($8.5$)-methyl acrylic acid (MAA).
After preparing the resist stack, the final pattern is written on the wafer.
The electron beam exposed wafer is then developed in a 1:3 solution of methyl isobutyl ketone (MIBK) and IPA.
After developing, the pattern of the PMMA layer is transferred to the germanium layer by reactive ion etching (RIE) with CF$_4$ gas.
 After this, an undercut is formed to the copolymer layer by oxygen plasma etching in the RIE machine.
 This last phase also removes any remaining PMMA. Now the final structure can be deposited through the holes in the Ge and copolymer layers.

To have different resistances for the detector and double dot, one needs to have individual control over the junction transparencies,
hence three angle shadow evaporation is used.
As the first step $14$~nm of Al (blue structures in Fig.~\ref{fig:SI1}; middle replica of the pattern) is evaporated at normal incidence.
Immediately following the deposition, without breaking the vacuum, the chip is exposed to 2 millibars of pure O$_2$ for 2 min for in-situ oxidation of the Al layer.
The oxidation is followed by
the evaporation of $30$~nm Cu (red structures in Fig.~\ref{fig:SI1}) at an angle 
so
 that Al from the first evaporation angle and Cu from this step form the junctions (with overlap area of the order of $50 \times 75 \; \rm nm^2 $) for both the right and left detectors.
The angles are adjusted so that the overlap in the detectors is not affecting that of the double dot, and vice versa.
The detector is evaporated first to ensure that it has lower resistance than the double dot, and to facilitate the measurement of electrons tunneling in the double dot.
Next pure O$_2$ at 5 millibars is used for further oxidation of Al layer.
As the final step,~$50$~nm of Cu (green structures in Fig.~\ref{fig:SI1}) is evaporated at an angle such that the overlap between this layer and the first Al layer forms the three
double dot junctions, each with an overlap area of $\approx 25 \times 50 \; \rm nm^2 $.

\section{Measurement setup} \label{a2}

The sample chip is enclosed in a sample stage~\cite{JPP_APL_10.1063}
with $12$ measurement lines and placed in a homemade dilution fridge with base temperature of $50$~mK.
All the signal lines are filtered by a Thermocoax cable
{with temperature} between 1 K and base temperature, and the sample stage
{is} thermally anchored to the mixing chamber.

The measurement setup used is shown schematically in Fig.~\ref{fig:SI1}.
 The bias voltages $V_{\text{b,det},\alpha}$, $\alpha=L,R$, across the leads of detectors (in blue)
 and $V_{\rm b}$, across the double dot leads (in blue and green), are applied using a commercial voltage source (Agilent 33522B).
 The DC gate voltages, $V_{g,\text{det},\alpha}$ and $V_{g,\alpha}$, tuning
 the offset charges
on the normal-metal islands (in red and green) and the superconducting island (blue) are
{also} applied using a commercial voltage source (Agilent 33522B).
The DC voltage signals are filtered with Thermocoax cables.
The single-electron currents $I_{\rm det,\alpha}^{\rm meas}(t)$ in the double dot are measured with a room-temperature current amplifier (Femto DLPCA-200).

During the measurement, the bias voltage ${V}_{\rm b}$ of the double dot is fixed to
a prescribed value.
The detector bias $V_{\rm {b,det,\alpha}}$($\alpha=L,R$ for the left and right detectors, respectively) and gate $V_{\rm {g,det,\alpha}}$ voltages are optimized to get the maximal signal to noise ratio. The backaction is not optimized leading to effective temperature of $ \approx 1$ K (see Sec.~\ref{Teff_cal}).

The output currents from left and right detectors, $I_{\text{det},L} $ and $I_{\text{det},R}$, are passed through two amplifiers
(DLPCA-200), with an amplification factor of $10^{10}$.
The amplifiers transform currents into voltage signals and the amplified signal of duration $15$~s is passed through an optoisolator and recorded by a 24-bit digitizer (NI 9239) at a sampling rate of $f$ = 25~kHz.
For each value of the bias voltage $V_{\rm b}$, we perform multiple measurements of duration $15$~s and combine the data into a single stationary trace of duration of the order of hours.

To characterize detectors and the double dot separately their $I$-$V$ characteristics are measured at different gate voltages $V_{g,\text{det},\alpha}$ and $V_{g,\alpha}$.
The resistances $R_{T,\text{det},L} = 1 ~\rm M\Omega$, $R_{T,\text{det},R} = 1.4 ~\rm M\Omega$ and charging energies $E_{C,\text{det},L}= 80 ~\mu \rm eV$, $E_{C,\text{det},R}= 90 ~\mu \rm eV$ of the detectors, and the common superconducting gap $\Delta = 200 ~\mu \rm eV$ are extracted from the $I$-$V$ characteristics using standard numerical simulations based on the Fermi's golden rule and the master equation~\cite{Kemppinen2009}.

To measure the $I$-$V$ characteristics of the double dot structure, we replace the grounding from the right end of the structure by an amplifier (Femto LCA-2-10T), with amplification coefficient $10^ {-12}$ A/V, connected to a digital multimeter (Agilent 34410A).
Both the gate voltages, $V_{\rm {g,L}}$ and $V_{\rm {g,R}}$ are swept for each value of the bias voltage $V_{\rm {b}}$.
For the lowest values of the bias voltage for which $|e V_{\rm {b}}|\lesssim 3\Delta$, a direct current measurement was not achievable due to its very low value ($\sim 10 ^{-18}$~A), thus the right end of the double dot was grounded as shown in Fig.~\ref{fig:SI1} and the output currents $I_{\rm det,\alpha}^{\rm meas}(t)$ from both the detectors were used to infer the current through the double dot.

 \begin{figure}
 \includegraphics[width= 0.5 \textwidth]{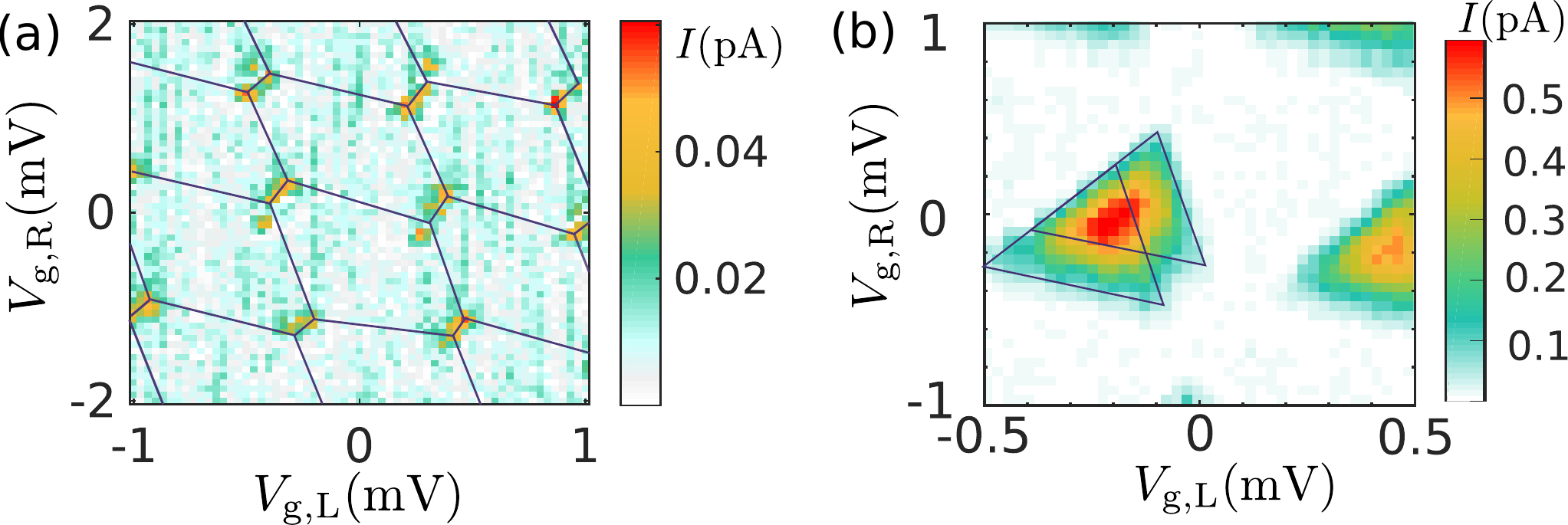}
\caption{ \textbf{Stability diagram of the double dot}.
Electric current across the double dot structure as a function of the two gate voltages $V_{g,L}$ and $V_{g,R}$ measured at high magnetic field, $H = 0.144 $~T, in which superconductivity of aluminum is suppressed.
(A)~Bias voltage, $V_{\rm b}$, is close to zero. A honeycomb pattern typical for double dot devices~\cite{RevModPhys.75.1} is visible. The high current  
spots correspond to the triple points at which the energies of three charge states are degenerate and the electric current can flow through the device.
(B)~$V_{\rm b} = 120 \;\mu\rm{V}$. The triple points grow into triangles because finite bias
allows electrons to pass through the double dot away from degeneracy.
The shape of the honeycomb structure and of the triangles allows us to determine the charging energies of the islands~\cite{RevModPhys.75.1}.
\label{fig:SI2}}
\end{figure}

The charging energies and resistances of the double dot are determined as follows:
\begin{enumerate}
\item A magnetic field $H = 0.144\,\rm T$ is applied to the superconducting part to increase the net current through the double dot. The applied magnetic field $H$ turns the superconductor into normal, thus increasing the number of electron tunneling events and the net current to $\approx 100$~fA.

\item We measure the current through the double dot for the bias voltage, $V_{\rm b} \approx 0 \;\mu\rm{V}$ and $V_{\rm b} = 120 \;\mu\rm{V}$, at different gate voltages $V_{g,L}$ and $V_{g,R}$
to obtain the stability diagram (see Fig.~\ref{fig:SI2}).
Comparing this diagram to the theory from \cite{RevModPhys.75.1} we extract the charging energies, $E_{C1} = 60\; \mu$eV, $E_{C,m} = 10\; \mu$eV and $E_{C2} = 40\; \mu$eV,
of left, middle and right double-dot junctions, respectively.

\item We obtain the total resistance of all three double-dot junctions in series to be
$R_L+R_m+R_R\simeq 55\;\rm M\Omega$, from room temperature $I$-$V$ measurement.
{Here, $R_L, \, R_M \, $ and $ \, R_R $ are the resistances of left, middle and right junctions of double dot structure, respectively.}

\end{enumerate}

\section{Double quantum dot: Charging energies, tunneling rates, detector back-action and effective temperature} \label{Teff_cal}

The theory of charge transport through a double dot is outlined in the review \cite{RevModPhys.75.1}.
In this section we use results relevant to our experiment
and adapt them to our particular setup, in which a double dot is capacitively coupled to two detectors.
We also clarify the mechanism of detector back-action, which leads to the enhanced effective temperature.

The Markovian dynamics of the system is governed by the master equation
\begin{eqnarray}
\dot P(n) = \sum_{m\not= n} \Gamma_{m}^n P(m) - \left(\sum_{m\not= n}\Gamma_{n}^m\right) P(n).
\label{master}
\end{eqnarray}
Here the indexes $m,n$ enumerate the four possible states of the double dot: $(0,0)$, $(1,0)$, $(0,1)$, and $(1,1)$; we use the shorthand notation
$\Gamma_{m}^n$ for the transition rate from the initial state $m$ to the final state $n$; and $P(n)$ is the
occupation probability of the state $n$.

The transition rates in Eq.~(\ref{master}) are determined by the resistances of three tunnel junctions, connecting
the dots and the leads, and by Coulomb energy barriers associated with electron tunneling. In order to determine
the latter we consider the energy of the whole system ``double quantum dot + detectors" (see Fig.~\ref{fig:1}A and Fig.~\ref{fig:SI1}),
\begin{eqnarray}
E(n) &=& E_{\rm dot}(n) + E_{C,{\rm det},L}(n_{{\rm det},L}-n_{g,{\rm det},L})^2
\nonumber\\ &&
+\, E_{C,{\rm det},R}(n_{{\rm det},R}-n_{g,{\rm det},R})^2
\nonumber\\ &&
+\, E_{c,L}(n_L - n_{g,L})(n_{{\rm det},L}-n_{g,{\rm det},L})
\nonumber\\ &&
+\, E_{c,R}(n_R - n_{g,R})(n_{{\rm det},R}-n_{g,{\rm det},R})
\nonumber\\ &&
+\, \frac{N_L-N_R}{2}eV_{\rm b}.
\label{Udot}
\end{eqnarray}
The first term in Eq.~\eqref{Udot}, $E_{\rm dot}(n)$,
is the electrostatic energy of the double dot, the second and the third terms are the
electrostatic energies of the detectors, the fourth and fifth terms describe the capacitive
coupling between the dots and the detectors, and the last term is the relevant part of the
energy of the voltage source. $N_L$ and $N_R$, appearing in the last term in~Eq.~\eqref{Udot}
 are the total numbers of electrons in the left and right leads, respectively.
For simplicity, we have omitted similar terms containing bias voltages applied to the detectors.
We have also assumed that the double dot is biased symmetrically, i.e. the potential of the left lead
is $V_{\rm b}/2$, while the potential of the right lead is $-V_{\rm b}/2$. This assumption is not restrictive since
any asymmetry in the bias may be absorbed in the shifts of gate voltages.
The energy of the double dot has the form
\begin{eqnarray}
E_{\rm dot}(n) &=& \frac{E_{C1}}{2}(n_L - n_{g,L})^2 + \frac{E_{C2}}{2}(n_R - n_{g,R})^2
\nonumber\\ &&
+\, E_{C,m}(n_L - n_{g,L})(n_R - n_{g,R}).
\label{E_ch_res}
\end{eqnarray}
The charging energies of the islands, $E_{C1},E_{C2}$,
of the detectors, $E_{C,{\rm det},L},E_{C,{\rm det},R}$, and the coupling energies $E_{C,m},E_{c,L},E_{c,R}$,
are defined as follows
\begin{eqnarray}
&& E_{C1}=\frac{e^2C_2}{C_0^2},\;
E_{C2}=\frac{e^2C_2}{C_0^2},
\nonumber\\ &&
E_{C,{\rm det},L} =\frac{e^2}{2C_{\Sigma,L}},\; E_{C,{\rm det},R} =\frac{e^2}{2C_{\Sigma,R}}, \;
E_{C,m}=\frac{e^2C_m}{C_0^2},
\nonumber\\ &&
E_{c,L}=\frac{2e^2C_{c,L}}{C_1C_{\Sigma,L}+\sqrt{C_1C_{\Sigma,L}(C_1C_{\Sigma,L}+4C_{c,L}^2)}},
\nonumber\\ &&
E_{c,R}=\frac{2e^2C_{c,R}}{C_2C_{\Sigma,R}+\sqrt{C_2C_{\Sigma,R}(C_2C_{\Sigma,R}+4C_{c,R}^2)}}.
\end{eqnarray}
The capacitances between different metallic electrodes of the system are defined in Fig.~\ref{fig:SI1};
the total capacitances of the islands read $C_1=C_L+C_{g,L}+C_m+\big[C_{c,L}^{-1}+(C_{{\rm det},L,1}+C_{{\rm det},L,2}+C_{g,{\rm det},L})^{-1}\big]^{-1}$,
$C_2=C_R+C_{g,R}+C_m+\big[C_{c,R}^{-1}+(C_{{\rm det},R,1}+C_{{\rm det},R,2}+C_{g,{\rm det},R})^{-1}\big]^{-1}$;
the total capacitances of the detectors are
$C_{\Sigma,L}=C_{{\rm det},L,1}+C_{{\rm det},L,2}+C_{g,{\rm det},L}+\big[C_{c,L}^{-1}+(C_{L}+C_{g,L}+C_{m})^{-1}\big]^{-1}$,
$C_{\Sigma,R}=C_{{\rm det},R,1}+C_{{\rm det},R,2}+C_{g,{\rm det},R}+\big[C_{c,R}^{-1}+(C_{R}+C_{g,R}+C_{m})^{-1}\big]^{-1}$;
the capacitance $C_0$ is defined as $C_0^2=C_1C_2-C_m^2$.
Here we have assumed that the capacitance between the two dots is small, $C_m\ll C_L,C_R$.
The dimensionless gate induced charges of the metallic islands read $n_{g,{\rm det},L}=C_{g,{\rm det},L}V_{g,{\rm det},L}/e$,
$n_{g,{\rm det},R}=C_{g,{\rm det},R}V_{g,{\rm det},R}/e$,
\begin{eqnarray}
n_{g,L} = \frac{C_{g,L}V_{g,L}}{e} + \frac{C_LV_{\rm b}}{2e}, \;
n_{g,R} = \frac{C_{g,R}V_{g,R}}{e} - \frac{C_RV_{\rm b}}{2e}.
\nonumber\\
\end{eqnarray}

Transitions between the charging states of the double dot occur if an electron
jumps through one of the three tunnel junctions. After a transition from the initial state $m$
to the final state $n$ the electron acquires
an energy $-Q_{m}^{n}=E(m)-E(n)$ which equals to the difference of the system energy Eq.~(\ref{Udot}) before
and after the jump. This energy gain is quickly redistributed between electrons, phonons etc., hence it can be
viewed as Joule heat associated with the transition. Minus sign in front of $Q_m^n$ comes from the convention used in the Main Text, where $Q$ is
considered to be positive if energy is extracted from the environment by the double dot.
The heat increments are antisymmetric, $Q_{m}^{n}=-Q^{m}_{n}$.
Therefore only six heat increments are needed to characterize the energetics of all 12 possible transitions in our system.
The corresponding heat exchanges evaluated at fixed values of the detector charges, which is indicated by the superscript $\sim$, read
\begin{eqnarray}
-\tilde Q^{00}_{10} &=& E_{C1}\left(\frac{1}{2} -n_{g,L}\right) - E_{C,m}n_{g,R} + \frac{eV_{\rm b}}{2}
\nonumber\\ &&
+\, E_{c,L}(n_{{\rm det},L}-n_{g,{\rm det},L}),
\nonumber
\end{eqnarray}
\begin{eqnarray}
-\tilde Q_{11}^{01} &=& E_{C1}\left(\frac{1}{2} -n_{g,L}\right) + E_{C,m}\left( 1 - n_{g,R} \right) + \frac{eV_{\rm b}}{2}
\nonumber\\ &&
+\, E_{c,L}(n_{{\rm det},L}-n_{g,{\rm det},L}),
\nonumber
\end{eqnarray}
\begin{eqnarray}
-\tilde Q_{10}^{11} &=& -E_{C,m}\left(1 - n_{g,L}\right) - E_{C2}\left(\frac{1}{2} - n_{g,R} \right) + \frac{eV_{\rm b}}{2}
\nonumber\\ &&
-\, E_{c,R}(n_{{\rm det},R}-n_{g,{\rm det},R}),
\nonumber
\end{eqnarray}
\begin{eqnarray}
 -\tilde Q^{01}_{00} &=& E_{C,m} n_{g,L} - E_{C2}\left(\frac{1}{2} - n_{g,R} \right) + \frac{eV_{\rm b}}{2}
\nonumber\\ &&
-\, E_{c,R}(n_{{\rm det},R}-n_{g,{\rm det},R}),
\nonumber
\end{eqnarray}
\begin{eqnarray}
-\tilde Q_{01}^{10} &=&
-(E_{C1}-E_{C,m})\left(\frac{1}{2} -n_{g,L}\right)
\nonumber\\ &&
+\, (E_{C2}-E_{C,m})\left( \frac{1}{2} - n_{g,R} \right)
\nonumber\\ &&
-\, E_{c,L}(n_{{\rm det},L}-n_{g,{\rm det},L})
\nonumber\\ &&
+\, E_{c,R}(n_{{\rm det},R}-n_{g,{\rm det},R}),
\nonumber
\end{eqnarray}
\begin{eqnarray}
-\tilde Q_{00}^{11} &=&
(E_{C1}+E_{C,m})\left(\frac{1}{2} -n_{g,L}\right)
\nonumber\\ &&
+\, (E_{C2}+E_{C,m})\left( \frac{1}{2} - n_{g,R} \right)
\nonumber\\ &&
-\, E_{c,L}(n_{{\rm det},L}-n_{g,{\rm det},L})
\nonumber\\ &&
+\, E_{c,R}(n_{{\rm det},R}-n_{g,{\rm det},R}).
\label{Qij}
\end{eqnarray}
These heat exchanges depend on instantaneous values of the charges of the detectors $n_{{\rm det},L}$ and $n_{{\rm det},R}$.
The latter fluctuate in time with typical frequency $I_{\rm det, \alpha} ^{\rm meas}/e\gtrsim 0.1$~GHz, which is much higher than the sampling
 data acquisition rate $f=25$~kHz. Hence experimentally measurable heat increments
are given by expressions (\ref{Qij}) averaged over the detector charges,
\begin{eqnarray}
Q_{m}^{n} = \left\langle \tilde Q_{m}^{n} \right\rangle_{n_{{\rm det},L},n_{{\rm det},R}}.
\label{Q_av}
\end{eqnarray}

The transition rate from the initial state $(m)$ to the final state $(n)$
at fixed $n_{{\rm det},L},n_{{\rm det},R}$ is given by
\begin{eqnarray}
\tilde\Gamma_{m}^{n} &=& \frac{1}{e^2 R_{nm}}\int dE N_i\left(E+\tilde Q_{m}^{n}\right)N_f(E)
\nonumber\\ && \times\,
f_i \left(E+\tilde Q_{m}^{n}\right)[1-f_f(E)].
\label{Gamma}
\end{eqnarray}
Here $R_{nm}$ is the resistance of the junction in which the electron jump occurs, $N_i(E)$ and $f_i(E)$ are, respectively, the density of states and distribution function
in the initial electrode, $N_f(E)$ and $f_f(E)$ are, respectively, the density of states and the distribution function in the destination electrode.
The density of states in the normal metals equals to~1, while in the superconductors it has the usual form
$N_S(E)=\theta(|E|-\Delta)|E|/\sqrt{E^2-\Delta^2}$, where $\Delta$ is the superconducting gap.
The transitions $(0,0)\leftrightarrow (1,1)$ occur by simultaneous cotunnelling of two electrons through two junctions.
The corresponding rates are defined by more complicated integrals, which we do not provide here for simplicity (for details of cotunnelling calculations in various Coulomb-blockaded systems see papers~\cite{averin1989macroscopic,averin1992single,konig1997cotunneling}).

According to our estimates, based on the measured transition rates in the interval $T_{\rm el} < Q_m^n < \Delta$,
where they almost do not depend on $Q_m^n$,
the distribution functions in all electrodes
can be rather well approximated by Fermi function with the electron temperature $T_{\rm el}\approx 170$~mK.
This temperature is higher than the base
temperature $50$~mK. The transition rates measured in the experiment are given by the integrals (\ref{Gamma})
averaged over the fluctuations of the detector charges,
\begin{eqnarray}
\Gamma_{m}^{n} = \left\langle \tilde \Gamma_{m}^{n} \right\rangle_{n_{{\rm det},L},n_{{\rm det},R}}.
\label{Gamma_av}
\end{eqnarray}
In Table I we list all experimentally measured rates $\Gamma_{m}^{n} $ for five   different values of the bias voltage $V_{\rm b}$.
Non-averaged rates satisfy the detailed balance condition
\begin{eqnarray}
{\tilde\Gamma_{m}^{n}}/{\tilde\Gamma^{m}_{n}} =
\exp\left[ {-\tilde Q_{m}^{n}}/{T_{\rm el}} \right].
\label{DB}
\end{eqnarray}
However, the detailed balance does not hold for the averaged rates $\Gamma_m^n$ (\ref{Gamma_av}) and average heat exchanges $Q_m^n$ (\ref{Q_av}) because it is broken
by back-action of the detectors: ${\Gamma_{m}^{n}}/{\Gamma^{m}_{n}} \neq
\exp\left[ {- Q_{m}^{n}}/{T_{\rm el}} \right]$.

The transport of electrons through the double dot occurs via two types of cyclic transitions between the charging states.
The cycle 1 involves the transitions $(0,0)\to(0,1)\to(1,0)\to(0,0)$, while the cycle 2 -- the transitions $(1,1)\to(0,1)\to(1,0)\to(1,1)$.
In both cases one electron is transferred from the right to the left lead.
 Thus, for sufficiently long observation time
the total charge transferred from the left lead to the right one reads
\begin{eqnarray}
q(t) = e N_1(t) + e N_2(t) + \delta q(t),
\label{Qt}
\end{eqnarray}
where we have used the fact that the electron charge is negative and equals to $-e$.
In Eq.~(\ref{Qt}) $N_1(t)=N_1^+ (t)-N_1^-(t)$ is the net number of completed cycles of the type 1 up to time $t$, i.e. the total number of completed cycles of type 1, $N_1^+ (t)$, minus the total number of cycles of type 1 completed in reverse order, $N_1^-(t)$. Similarly
 $N_2(t)$ and
$\delta q(t)$ denote, respectively, the net number of completed cycles of the type 2 and the contribution of incomplete cycles 1 or 2 up to time $t$.

An expression similar to Eq.~(\ref{Qt}) can be derived for stochastic entropy production. Namely, one finds
\begin{eqnarray}
S(t) = N_1(t)\mathcal{A}_1 +N_2(t)\mathcal{A}_2 + \delta S(t),
\label{St}
\end{eqnarray}
where $\mathcal{A}_1,\mathcal{A}_2$ are the affinities of the cycles 1 and 2 introduced before,
\begin{eqnarray}
\mathcal{A}_1= \log\frac{\Gamma_{00}^{01}}{\Gamma^{00}_{01}} \frac{\Gamma_{01}^{10}}{\Gamma^{01}_{10}} \frac{\Gamma_{10}^{00}}{\Gamma^{10}_{00}},\;
\mathcal{A}_2 =\log \frac{\Gamma_{11}^{01}}{\Gamma^{11}_{01}} \frac{\Gamma_{01}^{10}}{\Gamma^{01}_{10}} \frac{\Gamma^{11}_{10}}{\Gamma_{11}^{10}}.
\nonumber\\
\label{S12}
\end{eqnarray}
We have verified that for all bias voltages the cycles 1 and 2 give the dominating contribution to the entropy production.
The contribution of other cycles is suppressed by the low transition rates between the states $(0,0)$ and $(1,1)$.
Hence with a good accuracy we can omit the non-extensive terms $\delta q(t), \delta S(t)$ in Eqs.~(\ref{Qt}) and~(\ref{St}) in the long time limit,
when the contribution of incomplete cycles 1 or 2 also becomes small.

In the absence of detector back-action the detailed balance condition (\ref{DB}), in combination with the identities for the heat exchanges (\ref{Qij}), $\tilde Q_{00}^{01}+\tilde Q_{01}^{10}+\tilde Q_{10}^{00}=-eV_{\rm b}$, $\tilde Q_{11}^{01}+\tilde Q_{01}^{10}+\tilde Q_{10}^{11}=-eV_{\rm b}$,
imply that $\mathcal{A}_1=\mathcal{A}_2=eV_{\rm b}/T_{\rm el}$.
Comparing Eqs.~(\ref{Qt}) and (\ref{St}) we find that in this ideal case a simple relation between Joule heat and entropy production holds,
$\langle I\rangle V_{\rm b} = T_{\rm el}\langle\dot S\rangle$.
However, in the experiment detailed balance is broken by detector back-action.
Under these conditions the Joule heat and the entropy production are related via a proportionality constant,
\begin{eqnarray}
\langle I\rangle V_{\rm b} = T_{\rm eff}\langle\dot S\rangle,
\end{eqnarray}
defined as
\begin{eqnarray}
T_{\rm eff} \equiv \lim_{t\to\infty}\frac{q(t)V_{\rm b}}{S(t)}=
\frac{\langle N_1\rangle +\langle N_2\rangle }{\langle N_1\rangle \mathcal{A}_1+\langle N_2\rangle \mathcal{A}_2} eV_{\rm b}.
\label{Teff}
\end{eqnarray}

 \begin{table*}
 \begin{tabular}{|c|c|c|c|c|c|c|c|c|c|c|c| c|c|c|c|c|c|c|c|}
 \hline
 \multicolumn{1}{ |c| }{Bias voltages } & \multicolumn{12}{ |c| }{Transition rates (Hz) } & $T_{\rm eff}$ (K) & $\mathcal{A}_1$ & $\mathcal{A}_2$ \\
 \hline
 ($ \mu V$) & $\Gamma_{00}^{01}$ &$\Gamma_{01}^{00}$ &$\Gamma_{00}^{11}$ &$\Gamma_{11}^{00}$ &$\Gamma_{00}^{10}$ &$\Gamma_{10}^{00}$
& $\Gamma_{01}^{11}$ & $\Gamma_{11}^{01}$ & $\Gamma_{10}^{11}$ & $\Gamma_{11}^{10}$& $\Gamma_{01}^{10}$ & $\Gamma_{10}^{01}$ & & & \\
 \hline
 90.00 & 643.97 & 131.22 & 13.79 & 4.14 & 51.80 & 39.41 & 40.50 & 42.88 & 167.04 & 53.46 & 24.86 & 30.34 & 1.00 & 1.12 & 1.00
 \\
 \hline
 50.00 & 103.79 & 76.07 & 8.17 & 6.76 & 274.09 & 177.02 & 149.03 & 156.18 & 81.31 & 97.52 & 39.26 & 21.67 & 1.25 & 0.47 & 0.46
 \\
 \hline
 25.00 & 72.49 & 97.75 & 1.34 & 2.69 & 37.41 & 27.15 & 24.17 & 38.31 & 54.66 & 143.84 & 41.31 & 19.68 & 1.56 & 0.12 & 0.23
 \\
 \hline
 -25.00 & 90.89 & 81.84 & 1.64 & 2.21 & 36.21 & 27.64 & 25.86 & 39.85 & 67.40 & 113.97 & 25.13 & 31.66 & 0.79 & 0.40 & 0.32
 \\
 \hline
 -50.00 & 100.68 & 71.96 & 9.66 & 8.79 & 371.43 & 286.57 & 205.39 & 252.13 & 77.67 & 87.42 & 21.08 & 35.98 & 1.28 & 0.46 & 0.45
 \\
 \hline
	\end{tabular}
	\caption{\textbf{Transition rates between different charge states of the double dot, effective temperature and affinities in the cycles of the double dot for different values of the bias voltage.} The empirical transition rates are calculated using Eq.~(\ref{eq:Tr}). For $T_{\rm eff}$ we used Eqs.~(\ref{Teff}) and~(\ref{N12}) while for the cycle affinities $\mathcal{A}_{1,2}$ we used Eq.~(\ref{S12}).
	All the data is obtained from counting statistics of experimental traces of durations of at least 1 hour.}
	\label{tab:t1}
	\end{table*}

Applying usual full counting statistics methods~\cite{PhysRevB.67.085316}
to the master equation (\ref{master})
one can find the average numbers of cycles in the long-time limit, and under the assumption that $\Gamma_{00}^{11}=\Gamma_{11}^{00}=0$, $\langle N_1\rangle $ and $\langle N_2\rangle $ are given by
\begin{eqnarray}
\frac{\langle N_1\rangle}{t} &\simeq& \mathcal{N}_0^{-1}\big| \big(\Gamma_{00}^{01}\Gamma_{01}^{10}\Gamma_{10}^{00} - \Gamma^{00}_{01}\Gamma^{01}_{10}\Gamma^{10}_{00}\big)
\big( \Gamma_{11}^{10}+\Gamma_{11}^{01} \big)
\nonumber\\ &&
+\, \Gamma_{01}^{00}\Gamma_{11}^{01}\Gamma_{10}^{11}\Gamma_{00}^{10} - \Gamma^{01}_{00}\Gamma^{11}_{01}\Gamma^{10}_{11}\Gamma^{00}_{10} \big|,
\nonumber\\
\frac{\langle N_2\rangle}{t} &\simeq& \mathcal{N}_0^{-1} \big| \big( \Gamma_{11}^{01}\Gamma_{01}^{10}\Gamma^{11}_{10} - \Gamma^{11}_{01}\Gamma^{01}_{10}\Gamma_{11}^{10} \big)
\big( \Gamma_{00}^{10}+\Gamma_{00}^{01} \big)
\nonumber\\ &&
-\, \Gamma_{01}^{00}\Gamma_{11}^{01}\Gamma_{10}^{11}\Gamma_{00}^{10} + \Gamma^{01}_{00}\Gamma^{11}_{01}\Gamma^{10}_{11}\Gamma^{00}_{10} \big|.
\label{N12}
\end{eqnarray}
The normalization factor $\mathcal{N}_0$ is the same in both equations, it is the sum of various triple products of the rates.
Equations (\ref{Teff}) and~(\ref{N12}) fix the ratio $\langle N_1\rangle/\langle N_2\rangle$ and allow us to calculate $T_{\rm eff}$
counting cycles in the double dot experiment.
The values of $T_{\rm eff}$ obtained using Eq.~\eqref{N12} for different bias voltages are listed in Table~\ref{tab:t1}.
They vary from $0.79$~K to $1.56$~K with the average value around $1$~K, in agreement with the linear fit in
the inset of Fig.~\ref{fig:2}A in the Main Text. We also find that the cycle affinities $\mathcal{A}_1$ and $\mathcal{A}_2$ are rather close to each other
for all bias voltages except for $V_{\rm b}=25\mu\rm V$, where they differ by a factor of two, see Table~\ref{tab:t1}.

We now demonstrate that the value $T_{\rm eff}=1$K can be at least partially explained by the back-action of the detectors on the double dot.
Detailed analysis of back-action is not the main focus of this paper,
therefore we here restrict ourselves to simple estimates. We note that
averaging the rates (\ref{Gamma}) over detector charge fluctuations results in the replacement of the distribution function in the
normal metal by an effective distribution function. For example, if an electron jumps through the left junction, one should
replace the distribution function in the left normal dot by the following combination
\begin{eqnarray}
f_{\rm eff}^L(E) = \sum_{n_{{\rm det},L}} P_L(n_{{\rm det},L}) f_F\big(E - E_{c,L} \delta n_{{\rm det},L}, T_{\rm el} \big).
\nonumber
\end{eqnarray}
Here
\begin{eqnarray}
f_F(E,T) = 1/\big[1+e^{E/ T}\big]
\end{eqnarray}
is Fermi function, $\delta n_{{\rm det},L}=n_{{\rm det},L} - \langle n_{{\rm det},L} \rangle$,
$P_L(n_{{\rm det},L})$ is the probability for the left detector island to have $n_{{\rm det},L}$ extra electrons.
Next, one can roughly approximate the function $f_{\rm eff}^L(E)$ by a Fermi function with the same average energy of an electron,
$f_F(E,T_{\rm eff})$. Imposing the condition of equal average energies in the form
\begin{eqnarray}
\int \text{d}E\,E\, \big[ f_{\rm eff}^L(E) - f_F(E,T_{\rm eff}) \big] = 0,
\end{eqnarray}
we arrive at the following expression for the effective temperature of the left normal island
\begin{eqnarray}
T_{\rm eff}^L = \sqrt{T_{\rm el}^2 + \frac{12}{\pi^2}\eta_L^2 E_{C,{\rm det},L}^2\langle \delta n_{{\rm det},L}^2\rangle}.
\end{eqnarray}
Here we have defined the average squared fluctuations of the detector charge
$$
\langle \delta n_{{\rm det},L}^2\rangle = \sum_{n_{{\rm det},L}} P_L(n_{{\rm det},L})\, \delta n_{{\rm det},L}^2,
$$
and introduced the efficiency of the left detector $\eta_L = E_{c,L}/2E_{C,{\rm det},L}$. The latter is defined as
the shift of dimensionless gate charge of the detector induced by one extra electron in the left dot.
Repeating the same procedure, we find the effective temperatures of the normal leads adjacent to the middle and right junctions,
\begin{eqnarray}
T_{\rm eff}^m &=& \sqrt{T_{\rm el}^2 + \frac{12}{\pi^2}\sum_{s=L,R}\eta_s^2 E_{C,{\rm det},s}^2\langle \delta n_{{\rm det},s}^2 \rangle },
\\
T_{\rm eff}^R &=& \sqrt{T_{\rm el}^2 + \frac{12}{\pi^2}\eta_R^2 E_{C,{\rm det},R}^2\langle \delta n_{{\rm det},R}^2\rangle }.
\end{eqnarray}
In the experiment we find $\eta_L\approx \eta_R\approx 0.3$, $E_{C,{\rm det},L}=80$ $\mu$eV, $E_{C,{\rm det},R}=90$ $\mu$eV.
The detectors are biased a little bit above the conductance threshold $2(\Delta + E_{C,{\rm det, \alpha }})/e$. At this bias, and for high current state
of the detector, corresponding to one extra electron in the dot which the detector monitors, only
two allowed charging states of the detector island are populated, let's say $n_{\rm det, \alpha }=0$ and $n_{\rm det, \alpha }=1$.
Their occupation probabilities are approximately the same and equal to 1/2.
The average value of the detector charge then equals to $0.5$. Hence for both detectors we find
$\langle\delta n_{{\rm det},L}^2\rangle=\langle\delta n_{{\rm det},R}^2\rangle = [(0-0.5)^2+(1-0.5)^2]/2 = 0.25$.
With these parameters we find the following values of the effective temperatures
\begin{equation}
T_{\rm eff}^L = 0.23 \text{ K},\quad T_{\rm eff}^m = 0.29 \text{ K},\; \text{and}\quad T_{\rm eff}^R = 0.24\text{ K}.
\label{eq:leadTeff}
\end{equation}
These effective temperatures significantly exceed the electronic temperature of the double dot $T_{\rm el}\approx 170$~mK.
The value of $T_{\rm eff}$ for the whole device (\ref{Teff}) is even larger than the effective temperatures of the leads adjacent to the left, middle and right junctions given by~\eqref{eq:leadTeff}.
This is because $T_{\rm eff}$ is rather sensitive to the gate voltages $V_{g,L},V_{g,R}$. This dependence was ignored in our calculations.
Other back-action mechanisms, like, for example, emission of non-equilibrium phonons by the detectors may also contribute to increase $T_{\rm eff}$ and may require separate theoretical analysis.
Thus,
 our theoretical model reveals the significant contribution of the detector backaction to
the experimental value of $T_{\rm eff}$.

\section{Data Analysis} \label{ana}

 \begin{figure}
 \centering
 \includegraphics[width= 0.48 \textwidth]{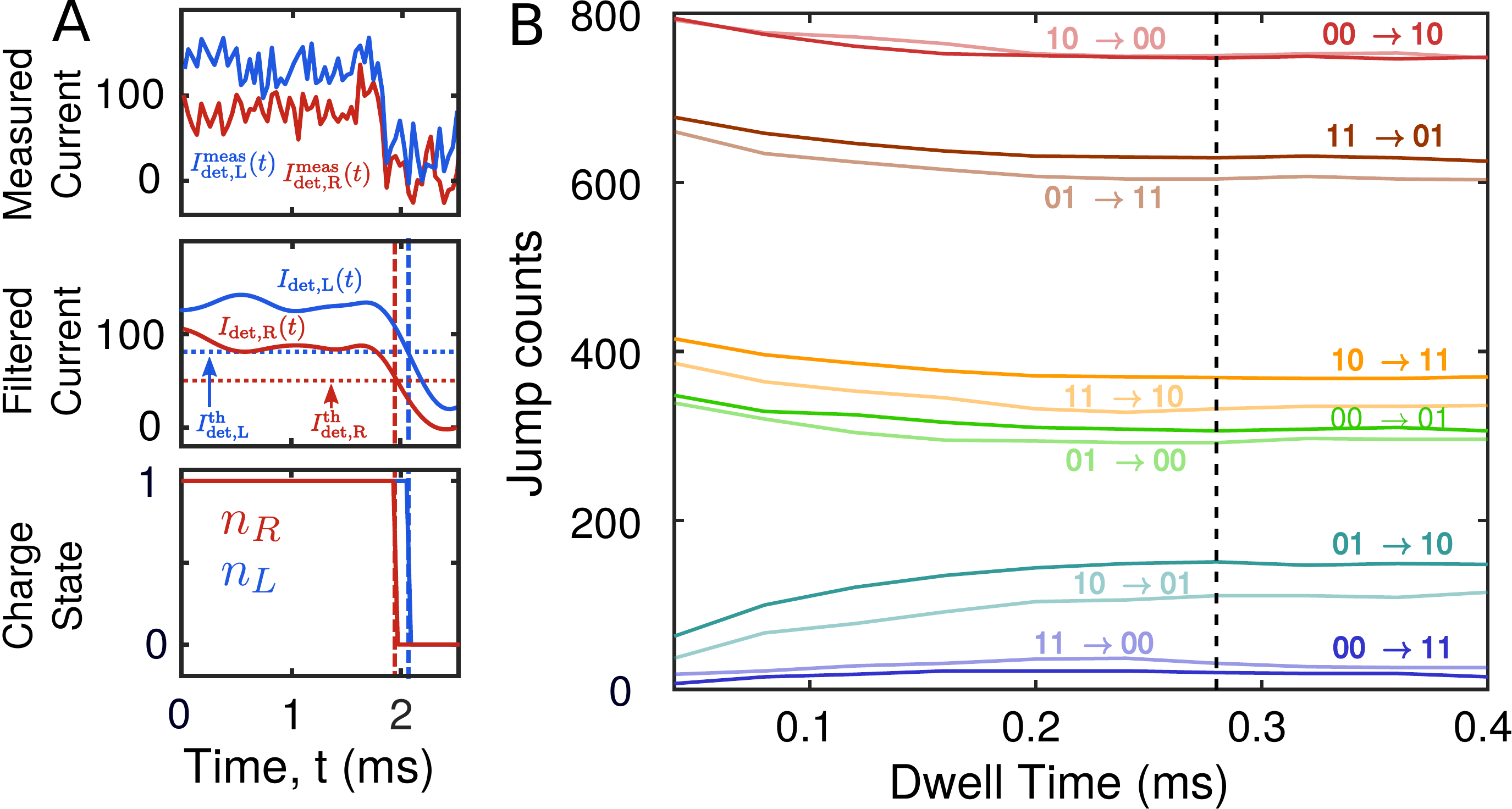}
\caption{ 
\textbf{ Filtering effect on current and use of dwell time as a corrective measure}.
(A)~Smearing of detector output currents introduced by filtering high
frequency noise. Top: measured detector currents, $I^{\rm meas} $,
for left detector in blue and right detector in red.
Middle:~the corresponding filtered current.
Bottom:~the corresponding charge state of the system island, $n$. 
Close to 2~ms both the detector currents jump instantaneously from higher state to the lower state in the measured current $I^{\rm meas} $
but after filtering, there is a delay introduced in the left detector current (blue curve) because of smearing of the transition point. 
(B)~Jump counts for each transition (see legend) as a function of the dwell time threshold used to correct for the stochastic jitter in the traces $\{\hat{n}(t)\}$ of the charge state of the double dot.
The dwell time threshold $\tau_{\rm th}$ is determined empirically as the minimal time threshold beyond which the jump counts for all transitions are barely affected upon small increments of the threshold.
For bias voltage $V_{\rm b} = 50 \; \mu$V shown in the figure, the dwell time threshold is estimated
to be 
$\tau_{\rm th} = 0.28 \; \rm ms$, illustrated by
the black vertical dashed line. 
Duration of the time trace analyzed here is 15 s.  
\label{fig:SI3}}
\end{figure}

The measured detector currents $I_{\rm det,\alpha}^{\rm meas}(t)$ ($\alpha=L,R$ for the left and right detectors, respectively) are filtered using a digital low-pass filter from MATLAB.
We use a fourth order infinite impulse response (IIR) low-pass filter with a cutoff frequency of $f_{\rm cut}=2$~kHz, because the changes in the detector signals, due to electron jumps, occur at a rate of $\sim$100~Hz.
We then discretize the filtered current $I_{\rm det,\alpha}(t)$ of each detector by assigning values $0$ or $1$ at each time as follows: i) first we compute the histogram of each detector current; ii) we introduce a current threshold $I_{\rm det,\alpha}^{\rm {th}}$ for each detector whose value is set at the local minimum between the two peaks of each histogram of the current  
($I^{\min}_\alpha$ and $I^{\max}_\alpha$); iii) we set the value of
 the charge state of the island $\alpha=L,R$ at time $t$, $n_\alpha (t)$, to the value $n_\alpha (t) = 1$ if the filtered current exceeds the threshold $I_{\rm det,\alpha}(t)>I^{\rm {th}}_{\rm det,\alpha}$ and we set $n_\alpha (t) =0$ if the filtered current is below the threshold value $I_{\rm det,\alpha}(t)<I^{\rm {th}}_{\rm det,\alpha}$. Such procedure is repeated systematically in each experiment for each detector current.
This is illustrated in Fig.~\ref{fig:SI3}A for a 2.5~ms time trace.

 \begin{figure}
 \centering
 \includegraphics[width= 80mm]{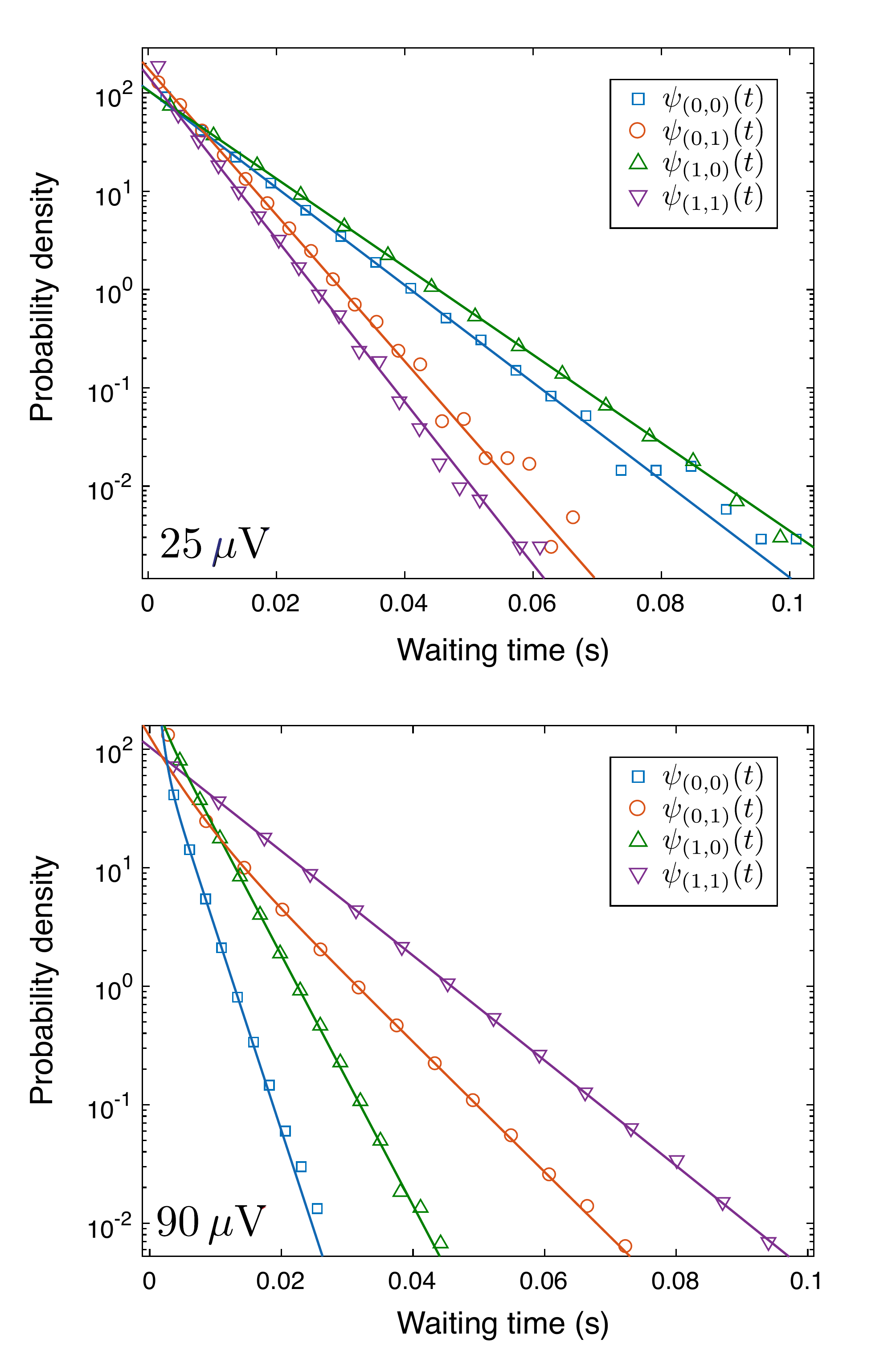}
\caption{\textbf{Experimental waiting-time distributions $\psi_{(n_L,n_R)}(t)$ for the different states $(n_L,n_R)$ in the double dot.}
Empirical waiting-time distributions (symbols) calculated from the traces $\{n_{\rm tot}(t)\}$ for $V_{\rm b}=25\mu\rm V$ (top) and $V_{\rm b}=90\mu\rm V$ (bottom). The solid lines are exponential fits.
}
\label{fig:wt}
\end{figure}

\begin{center}
		\begin{table*}
		\begin{tabular}{|c|c|c|c|c|c|c|}
		\hline
		\multicolumn{1}{ |c| }{Bias voltages } & \multicolumn{6}{ |c| }{Entropy jumps}
		\\
		\hline
		($ \mu V$) &
		$|S_{(0,0), (0,1)}| $ & $|S_{(0,0),(1,1)}| $ & $|S_{(0,0),(1,0)}| $ & $|S_{(0,1), (1,1)}|$ & $|S_{(1,0), (1,1)}| $ & $|S_{(0,1), (1,0)}| $
		\\
		\hline
		90.00 & 0.05 & 0.43 & 0.35 & 0.15 & 0.13 & 0.72
		\\
		\hline
		50.00 & 0.08 & 0.03 & 0.03 & 0.04 & 0.06 & 0.36
		\\
		\hline
		25.00 & 0.01 & 0.04 & 0.02 & 0.11 & 0.03 & 0.10
		\\
		\hline
		-25.00 & 0.05 & 0.07 & 0.13 & 0.08 & 0.03 & 0.21
		\\
		\hline
		-50.00 & 0.07 & 0.08 & 0.02 & 0.02 & 0.06 & 0.37
		\\
		\hline
	\end{tabular}
		\caption{\textbf{Entropy change corresponding to the jump between different charge state.}
		$|S_{n,n'}|$
		implies the absolute change in total entropy (Eq.~\eqref{eq:1})
		for the system jumping from state $n \rightarrow n'$.
		}
	\label{tab:t2}
	\end{table*}
\end{center}

Next, we combine $n_{\alpha}(t)$ into a trace $\hat{n}(t)=(n_{\rm L}(t),n_{\rm R}(t))$ describing the states of the double dot.
The low-pass filtering applied to the current signal $I_{\rm det, \alpha} ^{\rm meas}(t)$, introduces slight shifts of the time instants by the intervals of order of $1/f_{\rm cut}=0.5$~ms,
at which the jumps occur.
This jitter influences coincident jump events in both detectors
corresponding to the transitions $(0,0)\leftrightarrow(1,1)$ and $(0,1)\leftrightarrow(1,0)$.
An example of such an influence is illustrated in Fig.~\ref{fig:SI3}A. The two jumps in the detector currents,
which occur simultaneously (top panel), become slightly separated in time after the low-pass filtering (middle panel).
As a result, the apparent state trajectory evolves in time as $(1,1)\to (1,0) \to (0,0)$, see the lower panel.
The state $(1,0)$ is this sequence is clearly spurious, it did not exist in the original noisy signal shown in the top panel of Fig.~\ref{fig:SI3}A.
The above mentioned jitter can be compensated by the introduction of some ignorance in time shifts between signals of left and right detectors
 of order of $1/f_{\rm cut}=0.5$~ms.
Indeed, in order to eliminate these spurious states, we first identify the events in the low-pass-filtered time traces $\hat{n}(t)$
in which the two jumps in the left and right detectors occur close in time.
Next, we remove intermediate states between the initial and final states, for example $(1,1)$ and $(0,0)$,
if the dwell time in those states is shorter than a threshold value $\tau_{\rm th}$~\cite{chemla2008exact}.
After that, we treat the original two jumps as a single transition $(1,1)\to (0,0)$ occurring at time corresponding to the average value
between the times of the two jumps. We have varied the threshold time $\tau_{\rm th}$ and then taken the shortest value
above which the number of counts did not change significantly, see Fig.~\ref{fig:SI3}B. This value was found to be
$\tau_{\rm th} = 0.28$~ms in agreement with the estimated effect $1/f_{\rm cut}=0.5$~ms of the jitter.

To increase the statistics of jumps in the recorded traces we use the Markov properties of the state traces $\{n(t)\}$ and merge all $N$ $15$-second traces obtained for the same value of $V_{\rm b}$ into a single trace $\{n_{\rm tot}(t)\}$ of total duration $\tau=15N$ seconds.
This single trace $\{n_{\rm tot}(t)\}$ is used
both for calculation of stationary transition rates $\Gamma(n\to n')$ from state $n$ to state $n'$ and occupation probabilities $P^{\rm st}(n)$ of the state $n$
and for calculation of traces and statistics of the stochastic entropy production $S(t)$ and the entropy flow $S^e(t)$.

\begin{figure}
	\centering
	\includegraphics[width= 0.3\textwidth]{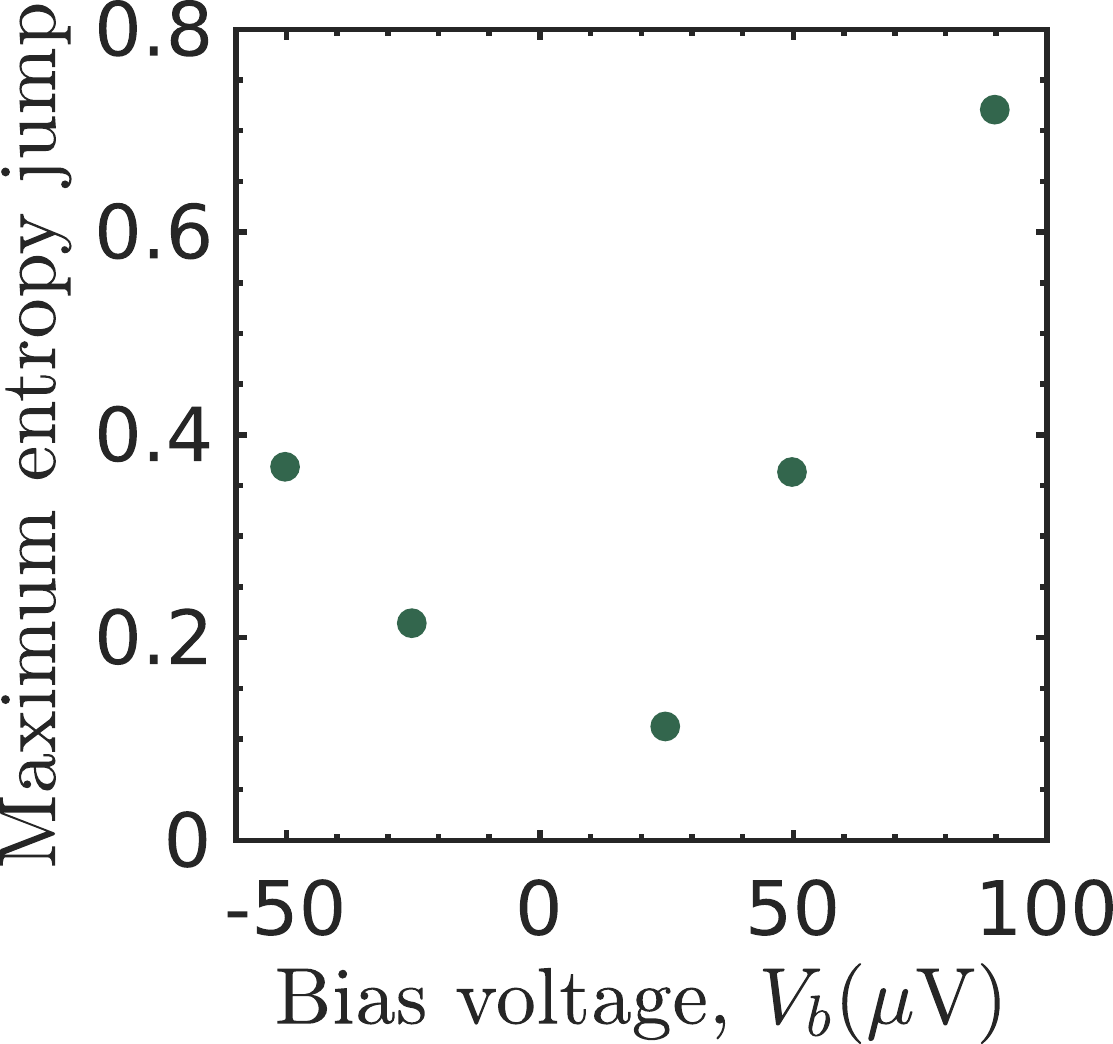}
	\caption{\textbf{Maximum entropy jump for different bias}.
		The maximum entropy change in a jump, $|S_{n,n'}|_{\rm max} = {\rm max}_{n,n'} |S_{n,n'}| $, as a function of bias voltage. Table~\ref{tab:t2} lists $|S_{n,n'}|$ for all possible jumps for different bias voltages. \looseness-1
		\label{fig:SI5}}
\end{figure}

To obtain the stationary transition rates $\Gamma(n\to n')$
from the time trace $\{n_{\rm tot}(t)\}$, we count the number of transitions $N_{n\to n'}$ that occur from state $n$ to state $n'$ for each bias voltage $V_{\rm b}$ value.
We calculate the transition rate between the states $n$ and $n'$ using \cite{Fujisawa1634}
\begin{equation} \label{eq:Tr}
\Gamma(n\to n')=\frac{N_{n\to n'}}{P^{\rm st}(n) \tau} \quad.
\end{equation}
where $\tau$ is the time duration of the experiment and
\begin{equation}
P^{\rm st}(n) = \tau_n/\tau,\label{eq:Pr}
\end{equation}
is the empirical steady-state occupation probability of the state $n$, calculated as the fraction of the total time when the double dot stays in state $n$. The traces of stochastic entropy production $S(t)$ and of the entropy flow $S^e(t)$ are calculated
using the empirical transitions rates~\eqref{eq:Tr}, occupation probabilities $P^{\rm st}_{n}$~\eqref{eq:Pr} from the time trace $\{n_{\rm tot}(t)\}$. The general formula for stochastic entropy production~\cite{seifert2012stochastic} is given by
\begin{equation}
S(t)= \log\frac{P(\{n(t)\})}{P(\{n(-t)\})},
\label{S_UDO}
\end{equation}
where $P(\{n(t)\})$ denotes the probability to observe the trajectory $\{n(t)\}$ and $P(\{n(-t)\})$ the probability to observe the corresponding time-reversed trajectory $\{n(-t)\}$. If the process $\{n(t)\}$ is Markovian, Eq.~\eqref{S_UDO} reduces to Eq.~(1) in the Main Text which we use for all our calculations of records of $S(t)$.
Figure~\ref{fig:wt} shows that the waiting-time distributions $\psi_{(n_L,n_R)}(t)$ for all the states of the double dot $(n_L,n_R)=\{(0,0),(0,1),(1,0),(1,1)\}$ are exponential for $V_{\rm b}=25\mu\rm V$ and $V_{\rm b}=90\mu\rm V$ thus confirming that the dynamics of the charge state of the double dot is indeed Markovian.
 The entropy jumps corresponding to transition between different system states for different bias voltages are listed in Table~\ref{tab:t2}.
 As seen here, the most dominant contribution to entropy is from electron jump between the island. In Fig.~\ref{fig:SI5}, we present the maximum jump in entropy resulting from the system state change for different bias voltage.

 To achieve an optimal statistical usage of the data for calculating entropy-production and the entropy flow to the environment records, we apply a sliding window procedure:
For a certain time window duration $\tau_{w}$ we use (possibly) overlapping sub-traces of $\{n_{\rm tot}(t)\}$
with time intervals $m\Delta t<t<m\Delta t +\tau_{w}$ for the $m$th sample trace.
The value $\Delta t$ of a time shift is chosen to be $10$ times larger than the decay time of the autocorrelation function of $\{n_{\rm tot}(t)\}$ to avoid unwanted correlations in different sample traces.

\section{Lower bound for the average negative record of the entropy flow} \label{a5}

We now demonstrate that the average negative record of the entropy flow is given by Eq.~(8) in the Main Text.

Equation~(1) of the main text can be rewritten as
\begin{equation}\label{eq:1_2}
S(t) = \log\frac{P^{\rm st}(n_0)}{P^{\rm st}(n_{\mathcal{N}(t)})} +S^e(t)\quad,
\end{equation}
where we have used the definition of the entropy flow $S^e(t)=\log \prod_{j=1}^{\mathcal{N}(t)} \Gamma (n_{j-1}\to n_j)/ \Gamma(n_j\to n_{j-1})$. Equation~\eqref{eq:1_2}
 and~Eq.~(7) of the Main Text imply
\begin{equation*}
\left\langle\min_{\tau\in[0,t]} \{ S^e(\tau) + \log P^{\rm st} (n_0) - \log P^{\rm st} (n_{\mathcal{N}(\tau)}) \}\right\rangle \geq -1.
\end{equation*}
Since
\begin{multline*}
\langle S^e_{\rm min}(t)\rangle +\sum_n P^{\rm st}(n) \log P^{\rm st}(n) -\log P^{\rm st}_{\rm min} \geq \\
\left\langle\min_{\tau\in[0,t]} \{S^e(\tau) + \log P^{\rm st} (n_0) - \log P^{\rm st} (n_{\mathcal{N}(\tau)}) \}\right\rangle ,
\end{multline*}
then Eq.~(8) of the Main Text follows. 


\begin{thebibliography}{73}
\bibitem [{nor()}]{nor_cur}%
	\BibitemOpen
	\href@noop {} {}\bibinfo {note} {The normalized detector currents
		($i_{\alpha}, \alpha=L R$) are defined as $ i_{\alpha}(t) = (I_{\rm
			det,\alpha}(t) - I^{\min}_\alpha)/(I^{\max}_\alpha -I^{\min}_\alpha)$.
		$I^{\max}_\alpha$ ($I^{\min}_\alpha$) is the detector current $I_{\rm
			det,\alpha} (t)$ corresponding to presence (absence) of an extra electron in
		the corresponding island}\BibitemShut {NoStop}%
\end{thebibliography}
%

\end{document}